\documentclass[usenatbib,a4paper]{mnras}

\usepackage{graphicx,threeparttablex,longtable,array} 
\usepackage{amssymb,amsmath,multirow,gensymb,lscape}
\usepackage{enumitem,multirow,subfig}
\usepackage{natbib}
\newcommand{\Jybeam}{\mbox{Jy beam$^{-1}$}}% mJy / beam
\newcommand{\mJyarcsec}{\mbox{mJy/arcsec$^{2}$/pix}}% mJy / beam
\newcommand{\fpeak}{\mbox{mJy/arcsec$^{2}$}}% mJy / beam
\newcommand{\um}{\mbox{$\mu$m}}% microns
\newcommand{\Msun}{\mbox{M$_{\odot}$}}

\newcommand{\CO}{\mbox{$^{12}$CO}}

\newcommand{\tOI}{\mbox{$t_{0}$}} %t_0I
\newcommand{\NIw}{\mbox{$N_{\textrm{I},w}$}} %NI,w
\newcommand{\NIt}{\mbox{$N_{\textrm{I},t}$}}%NI,t
\newcommand{\NI}{\mbox{$N_\textrm{I}$}}%NI
\newcommand{\tI}{\mbox{$t_\textrm{I}$}}%NI
\title[Embedded Binaries and Their Dense Cores]{Embedded Binaries and Their Dense Cores}

\author[Sadavoy \& Stahler]{Sarah I. Sadavoy$^{1,2}$
	 and Steven W. Stahler$^{3}$
	 \\
	 $^{1}$Max-Planck-Institut f\"{u}r Astronomie (MPIA), K\"{o}nigstuhl 17, D-69117 Heidelberg, Germany\\
	 $^{2}$Harvard-Smithsonian Center for Astrophysics, 60 Garden Street, Cambridge, MA 02138, USA\\
	 $^{3}$Astronomy Department, University of California, Berkeley, CA 94720, USA
	 }

\begin{document}

% These dates will be filled out by the publisher
\date{Accepted 2017 April 28. Received 2017 April 25; in original form 2017 February 28}

% Enter the current year, for the copyright statements etc.
\pubyear{2017}

\label{firstpage}
\pagerange{\pageref{firstpage}--\pageref{lastpage}}
\maketitle

%%%%%%%%%%%   ABSTRACT  %%%%%%%%%%%%%%%%
\begin{abstract}
We explore the relationship between young, embedded binaries and their parent cores, using observations within the Perseus Molecular Cloud.  We combine recently published VLA observations of young stars with core properties obtained from SCUBA-2 observations at 850 \um.   Most embedded binary systems are found toward the centres of their parent cores, although several systems have components closer to the core edge.  Wide binaries, defined as those systems with physical separations greater than 500 au, show a tendency to be aligned with the long axes of their parent cores, whereas tight binaries show no preferred orientation.  We test a number of simple, evolutionary models to account for the observed populations of Class 0 and I sources, both single and binary.  In the model that best explains the observations, all stars form initially as wide binaries.  These binaries either break up into separate stars or else shrink into tighter orbits.  Under the assumption that both stars remain embedded following binary breakup, we find a total star formation rate of 168 Myr$^{-1}$.  Alternatively, one star may be ejected from the dense core due to binary breakup.  This latter assumption results in a star formation rate of 247 Myr$^{-1}$.  Both production rates are in satisfactory agreement with current estimates from other studies of Perseus.  Future observations should be able to distinguish between these two possibilities.  If our model continues to provide a good fit to other star-forming regions, then the mass fraction of dense cores that becomes stars is double what is currently believed.
\end{abstract}


%\keywords{keywords go here}

%%%%%%%%%%%%%%%%%%%%%%%%%%%%%%%%%%%%%
%%%%%%%%%%%   INTRODUCTION  %%%%%%%%%%%%%%%%
%%%%%%%%%%%%%%%%%%%%%%%%%%%%%%%%%%%%%
\section{Introduction\label{Intro}}

The origin of binary stars has long been one of the central problems of astronomy.  Populations of multiple star systems at the main sequence and pre-main sequence stages have been well studied in various clouds \citep[e.g.,][]{Duquennoy91, Raghavan10, Kraus11}, revealing variations in binary frequency with stellar mass \citep{Lada06, DucheneKraus13}.  The binarity at these evolutionary phases, however, has been influenced by stellar dynamical interactions and does not represent the primordial binary distribution \citep[e.g.,][]{MarksKroupa12, Reipurth14}.  Observers have also searched for even younger binaries whose components are Class 0 or Class I sources, which are still embedded within dusty clouds \citep[e.g.,][]{LadaWilking84, Andre93, Greene94}.  

Early studies of embedded multiple systems suggested higher binary fractions at younger stages than the main sequence phase \citep[e.g.,][]{Reipurth00, Looney00, Haisch04, Duchene04}.  These investigations, however, made no pretense of completeness.  Several recent studies have improved the statistics to better address Class 0 and Class I binary fractions.  \citet{Connelley08} observed over 200 Class I sources within a number of star-forming regions in the near-infrared, establishing the first robust binary fractions for mostly wide separations at this stage \citep{Duchene07, Connelley09}.  Analogous studies of Class 0 objects have been published using millimeter interferometry observations, but with smaller numbers of targets \citep{Maury10, Chen13}.  More recently, \citet{Tobin16} used the VLA to conduct a complete study of all Class 0 and I sources in the Perseus molecular cloud (the VANDAM survey), producing the largest, uniform study of multiplicity and spatial separations in embedded binaries.

To advance our understanding of these embedded systems further, we should investigate the relationship between the young binary systems and their gaseous environments.  It has long been accepted that both Class 0 and I sources are embedded within dense cores inside larger molecular cloud complexes \citep[e.g., see ][for review]{difran07, Andre14}.  While much work has been made in establishing populations of embedded binaries and their intrinsic separation distributions, the basic properties of their parent dense cores and the locations of the stellar components within them remain unclear.   Connecting the stars to their dense cores will provide new insights into how the binaries themselves originate and evolve.

In this paper, we combine the binary database from the VANDAM survey \citep{Tobin16} with the associated dense cores, identified from SCUBA-2 observations for the Perseus molecular cloud \citep[$d=235$ pc;][]{Hirota08}.  In Section \ref{binaries}, we present the binary systems, while Section \ref{cores} identifies the associated dense cores and lists their basic properties.  In Section \ref{results}, we discuss the spatial relation of the binaries to their host cores.  We find that most relatively wide binaries, with separations exceeding 500 au, tend to be aligned with the long axis of their parent cores.  In contrast, the relatively tight binaries, with separations smaller than 500 au, have no preferred alignment.  In Section \ref{theory}, we test simple evolutionary models for the observed populations of embedded systems, both single and binary.  We show that the observed populations are best recovered when all stars form initially as wide binaries.  Finally, in Section \ref{summary}, we summarize our results and propose future studies that bear on the question of binary origin.

%%%%%%%%%%%%%%%%%%%%%%%%%%%%%%%%%%%%%%%%
%%%%%%%%%%%%%%  The DATA section  %%%%%%%%%%%%%%%%
%%%%%%%%%%%%%%%%%%%%%%%%%%%%%%%%%%%%%%%%

\section{Identifying the Binaries}\label{binaries}

We use the source catalogue from the VANDAM survey \citep{Tobin16} to identify all the embedded sources in Perseus down to separations of $\sim$ 20 au.  In brief, the VANDAM survey targeted 93 objects in Perseus, including all known embedded sources, with the VLA in the Ka radio band ($\sim 9$ mm).  The observations conducted single pointings with the A- and B-array configurations, reaching resolutions of $\sim$ 15 au.  The initial source catalogue used detections at $\sim$ 9 mm with emission at SNR $\gtrsim 5$, resulting in over 100 compact radio sources.  Although the VANDAM survey also included more evolved sources, we use only those sources identified as Class 0 or Class I in our analysis based on the source classifications given in \citet{Tobin16}. 

In total, we include 71 embedded sources from the VANDAM survey in our analysis.   Not all of these embedded objects are in multiple systems, however.  \citet{Tobin16} used a conservative upper limit separation of 10$^4$ au to identify 26 multiple systems.  Here, we identify binaries and higher-order multiples as systems with more than one stellar member associated with the same dense core (see Section \ref{coreID}).   

\section{Identifying the Dense Cores} \label{cores}

\subsection{SCUBA-2 Observations}

We use SCUBA-2 observations at 850 \um\ from the JCMT Gould Belt Survey \citep{Ward-T07} to identify the dense cores.  These data were published and fully described in \citet{MChen16} and the data reduction is described in \citet{Mairs15}.  Perseus was observed by SCUBA-2 at 450 \um\ and 850 \um\ in ten $\sim$ 30\arcmin\ fields.  Each field was observed multiple times with a PONG1800 observing pattern, where the instrument makes five square scans that are successively rotated by 18\degree\ on the sky \citep[see][]{Kackley10, Holland13, Bintley14}.  Each field was reduced separately and then mosaicked together using the SMURF package in \emph{Starlink} \citep[][]{Jenness11, Chapin13, Currie14} using user-defined masks to identify regions of bright emission in a second iteration of the reduction \citep[e.g., see][]{Mairs15}.

The data presented here corresponds to Data Release 1 (DR1).  For all DR1 observations, the 850 \um\ emission was automatically corrected for \CO\ (3-2) flux contamination \citep[e.g., see][]{Drabek12} for those regions that have complementary \CO\ (3-2) observations from HARP \citep[e.g., NGC1333, IC348, L1448, L1455, B1;][]{Buckle09, Curtis10, Sadavoy13}.   The final map has a sensitivity of $\sim$ 30 m\Jybeam\ \citep{MChen16}.  We adopt an effective beam of 14.6\arcsec\ at 850 \um\ based on a two-component beam fit that includes the error beam \citep{Dempsey13}.

\subsection{Core Identification}\label{coreID}

We use the source extraction algorithm, \emph{getsources} \citep{getsources, getsources+filaments} to identify the dense cores in the SCUBA-2 850 \um\ data.  In brief, \emph{getsources} runs a set of spatial decompositions over various spatial scales to identify cores and filaments from intensity moments in the filtered images.  Extractions from the different spatial decompositions are used to build a deblended source catalogue and to characterize a local background level  for each source.  The final source properties (flux, size, orientation) are determined from the intensity moments with this local background removed.  

\begin{figure}
\includegraphics[width=\columnwidth]{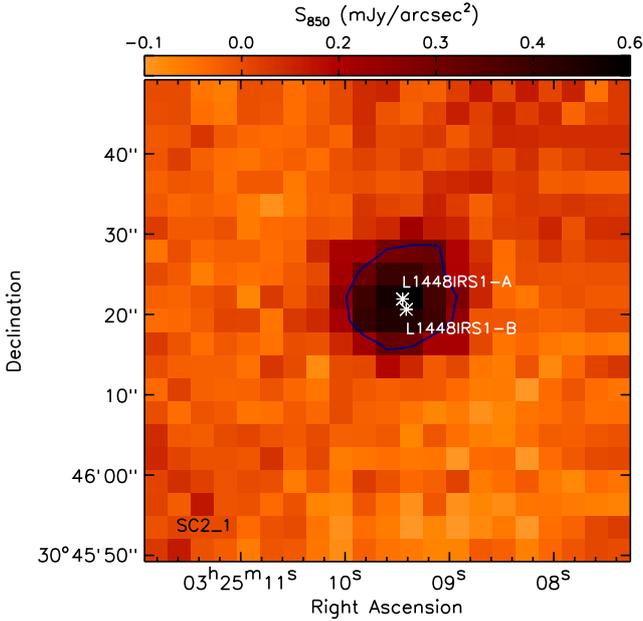}
\caption{SCUBA-2 850 \um\ map of SC2\_1 (L1448 IRS1).  The map is $\sim 1$\arcmin\ on a side, centered at the position of the core (see Table \ref{coreTable}).  The associated embedded sources from \citet{Tobin16} are shown with white stars and labels.  Blue contours show a flux density of 0.3 \mJyarcsec, which corresponds to 50\%\ of the peak core flux. Maps of all cores with embedded binaries are given in Appendix \ref{appendix}.\label{example_map}}
\end{figure}

Although we have source extractions for the complete SCUBA-2 850 \um\ dataset of Perseus, we focus only on those cores associated with multiple systems in this study\footnote{Complete catalogues of the Perseus 850 \um\ SCUBA-2 observations will be provided by the SCUBA-2 Gould Belt Survey team.  Maps from \citet{MChen16} are available here: https://doi.org/10.11570/16.0004}.  We associate the VLA-identified stellar sources with a core if (1) they are within the 50\%\ flux contour of that core and (2) they are within the full width at half max (FWHM) of the core as determined by \emph{getsources}.  Using these criteria, we find 24 embedded multiple systems with a total of 55 embedded sources.  Figure \ref{example_map} shows an example of one core (labeled here as SC2\_1) with its associated binary system.  We show maps for the full catalogue of stellar systems and their cores in Appendix \ref{appendix}.

Table \ref{coreBinary} lists the embedded binaries and their associated cores.  The first and second columns give the stellar source designation and stellar classification of the spectral energy distribution from \citet{Tobin16}.  The third column gives the distance between each embedded source and its parent core center.  Finally, in the last two columns, we give the IAU designations based on the core center and shorter core numbers that we will use in the text.    For simplicity, we consider the stellar components identified as Class 0/I to be Class I objects and all components identified as first hydrostatic cores \citep[e.g.,][]{Larson69, Pezzuto12} to be Class 0 sources.   We also assume all sources within the same core are at the same evolutionary stage (see Section \ref{noncoeval}).

{\setlength{\extrarowheight}{0.4pt}%
\begin{table*}
\renewcommand{\TPTminimum}{\linewidth}
\caption{Embedded Binary Systems}\label{coreBinary}
\begin{threeparttable}
\begin{tabular}{lccll}
\hline\hline
Source$^{a}$	& Class$^{a}$	&  D$^{b}$ (au)	& Core IAU Designation	&  Core Number \\
%Core	 &  RA (J2000) & Dec(J2000) &	YSOs	& D\tablefootmark{b} (AU)	 \\
\hline
L1448IRS1-A 		& I 		& 94 		& JCMTLSG J032509.5+304622	& SC2\_1 	  \\
L1448IRS1-B 		& I 		& 287 	& $\cdots$ & $\cdots$ 		\\
\hline
Per-emb-22-B 		& 0 		& 63	 	& JCMTLSG J032522.3+304513 	& SC2\_2 	 \\
Per-emb-22-A 		& 0 		& 202	& $\cdots$ & $\cdots$ 		 \\
\hline
L1448NW-B 		& 0 		& 924 	& JCMTLSG J032535.6+304538 	& SC2\_3 	 \\
L1448NW-A 		& 0 		& 977 	& $\cdots$ & $\cdots$ 		 \\
\hline
Per-emb-33-B 		& 0 		& 163 	& JCMTLSG J032536.3+304516 	& SC2\_4 	 \\
Per-emb-33-C 		& 0 		& 217 	& $\cdots$ & $\cdots$ 		 \\
Per-emb-33-A 		& 0 		& 320 	& $\cdots$ & $\cdots$ 		\\
L1448IRS3A 		& I 		& 1520 	& $\cdots$ & $\cdots$ 		 \\
\hline
Per-emb-26 		& 0 		& 397 	& JCMTLSG J032538.9+304404 	& SC2\_5 	 \\
Per-emb-42 		& I 		& 1597 	& $\cdots$ & $\cdots$ 		 \\
\hline
Per-emb-48-A 		& I 		& 405 	& JCMTLSG J032738.4+301358 	& SC2\_6 	 \\
Per-emb-48-B 		& I 		& 465 	& $\cdots$ & $\cdots$ 		 \\
\hline
Per-emb-17-B		& 0 		& 356 	& JCMTLSG J032739.2+301303 	& SC2\_7 	 \\
Per-emb-17-A 		& 0 		& 402 	& $\cdots$ & $\cdots$ 		 \\
\hline
Per-emb-35-A 		& I 		& 328 	& JCMTLSG J032837.1+311332 	& SC2\_8 	 \\
Per-emb-35-B 		& I 		& 485 	& $\cdots$ & $\cdots$ 		 \\
\hline
Per-emb-27-A 		& 0/I 		& 298 	& JCMTLSG J032855.6+311438 	& SC2\_9 	 \\
Per-emb-27-B 		& 0/I 		& 433 	& $\cdots$ & $\cdots$ 		 \\
\hline
Per-emb-36-A 		& I 		& 123 	& JCMTLSG J032857.4+311416 	& SC2\_10 \\
Per-emb-36-B 		& I 		& 133 	& $\cdots$ & $\cdots$ 		 \\
\hline
SVS13A2 			& 0/I 		& 488 	& JCMTLSG J032903.4+311600 	& SC2\_11 \\   %Tobin considers SVS13A 0/I
Per-emb-44-B 		& 0/I 		& 1399 	& $\cdots$ & $\cdots$ 		 \\
Per-emb-44-A 		& 0/I 		& 1455 	& $\cdots$ & $\cdots$ 		 \\
SVS13B 			& 0 		& 2102 	& $\cdots$ & $\cdots$ 		 \\
\hline
Per-emb-12-A 		& 0 		& 446 	& JCMTLSG J032910.6+311333 	& SC2\_12 \\
Per-emb-12-B 		& 0 		& 451 	& $\cdots$ & $\cdots$ 		 \\
\hline
Per-emb-18-A 		& 0 		& 865 	& JCMTLSG J032911.2+311828 	& SC2\_13 \\
Per-emb-18-B 		& 0 		& 873 	& $\cdots$ & $\cdots$ 		 \\
Per-emb-21 		& 0 		& 2285 	& $\cdots$ & $\cdots$ 		 \\
\hline
Per-emb-13 		& 0 		& 489 	& JCMTLSG J032912.2+311309 	& SC2\_14 \\
IRAS4B$^{\prime}$ 	& 0 		& 2123	& $\cdots$ & $\cdots$ 		 \\
\hline
Per-emb-49-B 		& I 		& 2438 	& JCMTLSG J032913.7+311810 	& SC2\_15 \\
Per-emb-49-A 		& I 		& 2489 	& $\cdots$ & $\cdots$ 		 \\
\hline
Per-emb-37 		& 0 		& 251 	& JCMTLSG J032919.0+312315 	& SC2\_16 \\
EDJ2009-235 		& II 		& 2476 	& $\cdots$ & $\cdots$ 		 \\
\hline
Per-emb-5-B 		& 0 		& 1464 	& JCMTLSG J033120.6+304526	& SC2\_17 \\
Per-emb-5-A 		& 0 		& 1482 	& $\cdots$ & $\cdots$ 		\\
\hline
Per-emb-2-A 		& 0 		& 817 	& JCMTLSG J033217.7+304946 	& SC2\_18 \\
Per-emb-2-B 		& 0 		& 808 	& $\cdots$ & $\cdots$ 		 \\
\hline
Per-emb-40-A 		& I 		& 1909 	& JCMTLSG J033316.1+310752 	& SC2\_19 \\
Per-emb-40-B 		& I 		& 1970 	& $\cdots$ & $\cdots$ 		 \\
\hline
B1-bS$^{d}$ & FHSC$^{d}$  & 780 		& JCMTLSG J033321.3+310729 	& SC2\_20\\
B1-bN$^{d}$ & FHSC$^{d}$  & 3343 	& $\cdots$ & $\cdots$ 		 \\
\hline
Per-emb-16 		& 0 		& 664 	& JCMTLSG J034351.2+320322 	& SC2\_21 \\
Per-emb-28		& 0 		& 3400 	& $\cdots$ & $\cdots$ 		 \\
\hline
Per-emb-11-A 		& 0 		& 544 	& JCMTLSG J034357.2+320305 	& SC2\_22 \\
Per-emb-11-B 		& 0 		& 1206 	& $\cdots$ & $\cdots$ 		 \\
Per-emb-11-C 		& 0 		& 1733 	& $\cdots$ & $\cdots$ 		\\
\hline
Per-emb-32-B 		& 0 		& 3730 	& JCMTLSG J034401.5+320153 	& SC2\_23 \\
Per-emb-32-A 		& 0 		& 3889 	& $\cdots$ & $\cdots$ 		 \\
\hline
Per-emb-8 		& 0 		& 364 	& JCMTLSG J034444.1+320134 	& SC2\_24 \\
Per-emb-55-B 		& I 		& 2262  	& $\cdots$ & $\cdots$ 		 \\
Per-emb-55-A 		& I 		& 2392	& $\cdots$ & $\cdots$ 		 \\
\hline
\end{tabular}
\begin{tablenotes}[normal,flushleft]
\item \tnote{a} Taken from \citet{Tobin16}.  \tnote{b} Separation between each embedded source and the core center (given by the RA/Dec coordinates) assuming a distance of 235 pc  \citep{Hirota08}.  Stars are ordered by increasing distance from the core center (see Table \ref{coreTable}).  \tnote{c}  For brevity, we refer to the cores by their running number rather than their full IAU name. \tnote{d} B1-bS and B1-bN binary are identified as two distinct objects with \emph{getsources}, although they share a common envelope (see text).  We consider B1-bS and B1-bN to be part of the same system and use the core centered on B1-bS for our analysis.  These objects are classified as first hydrostatic cores (FHSC) in \citet{Tobin16}.
\end{tablenotes}
\end{threeparttable}
\end{table*}	
}

We note that several stellar systems from \citet{Tobin16} break up into separate systems based on their associations with different SCUBA-2 identified cores.   In these cases, we consider the two ``cores'' to be in the same systems if they share a common envelope.  For simplicity, we identify two nearby objects as having a common envelope if they overlap (based on their getsources extractions) and lie within their respective 50\%\ flux contours.  For example, SC2\_3 and SC2\_4 (see Figure \ref{fig:allfeature1}) are two cores associated with the L1448 IRS3 system, a well-known, young multiple system.  Since SC2\_3 is much fainter than SC2\_4 (by a factor of three) and lies outside of its 50\%\ flux contour, we consider these two cores (and their corresponding embedded sources) to be distinct multiple systems.  In contrast, we consider the two cores seen toward SC2\_20 (see Figure \ref{fig:allfeature1}) in the B1-b system to be part of the same system, because they each lie within the 50\%\ contour of the other.  Indeed, if we subtract the \emph{getsources} objects from the continuum map, we find a smooth large-scale envelope at 850 \um.  

Table \ref{coreTable} lists the properties of each core from the \emph{getsources} extractions, including position, peak 850 \um\ flux, total 850 \um\ flux, semi-major axis ($a$), semi-minor axis ($b$), and position angle ($\theta$).  Many of the cores are compact (e.g., unresolved or partially resolved with a 14.6\arcsec\ beam), although several appear large and extended (see also, Figure \ref{fig:allfeature1}).  

\begin{table*}
\begin{threeparttable}
\caption{Dense Cores Associated with Embedded Multiples}\label{coreTable}
\begin{tabular}{lccccccc}
\hline\hline
Core	Number & RA			&	Dec	       & $S_{peak}$$^{a}$ & $S_{total}$$^{a}$	&  $a$$^{a}$	&   $b$$^{a}$	&  $\theta$$^{a}$\\
	&(J2000)		&	(J2000)   & 		  (\fpeak) 	 &	(Jy)	&(arcsec) & (arcsec)	& (deg) \\
\hline
SC2\_1 	& 3:25:09.5 & 30:46:22 &     0.59 	&   0.11 	&  14.6 &  14.6 &   131  \\  
SC2\_2 	& 3:25:22.3 & 30:45:13 &     4.24 	&   1.30 	&  14.6 &  14.6 &   72  \\	
SC2\_3 	& 3:25:35.6 & 30:45:38 &    5.39 	&   1.65 	&  14.6 &  14.6 &   55  \\
SC2\_4	& 3:25:36.3 & 30:45:16 &    15.6 	&   4.61 	&  14.6 &  14.6 &   14  \\	
SC2\_5   	& 3:25:38.9 & 30:44:04 &     6.26 	&   2.01 	&  14.6 &  14.6 &   139  \\	
SC2\_6 	& 3:27:38.4 & 30:13:58 &     0.73 	&   0.51 	&  27.0 &  16.9 &   28  \\	
SC2\_7 	& 3:27:39.2 & 30:13:03 &     2.07 	&   0.67 	&  15.0 &  14.6 &   167  \\	
SC2\_8 	& 3:28:37.1 & 31:13:32 &     1.95 	&   0.69 	&  15.2 &  14.6 &   51  \\	
SC2\_9   	& 3:28:55.6 & 31:14:38 &     9.91 	&   2.69 	&  14.6 &  14.6 &   156  \\	
SC2\_10 	& 3:28:57.4 & 31:14:16 &     1.88 	&   2.16 	&  15.8 &  14.6 &   94  \\	
SC2\_11   & 3:29:03.4 & 31:16:00 &     10.28  	&  4.52 	&  20.7 &  14.6 &   33  \\	
SC2\_12   & 3:29:10.6 & 31:13:33 &    33.99 	&  8.07 	&  14.6 &  14.6 &   163  \\	
SC2\_13 	& 3:29:11.2 & 31:18:28 &    4.04 	&   1.98 	&  22.1 &  14.6 &   26  \\	
SC2\_14   & 3:29:12.2 & 31:13:09 &    13.62 	&  4.04 	&  14.7 &  14.6 &   105  \\	
SC2\_15 	& 3:29:13.7 & 31:18:10 &     0.81 	&   0.39	&  21.4 &  16.7 &   138  \\	 
SC2\_16 	& 3:29:19.0 & 31:23:15 &    1.33 	&   0.49 	&  16.9 &  14.6 &   161  \\	
SC2\_17 	& 3:31:20.6 & 30:45:26 &    2.45 	&   1.33 	&  19.8 &  14.6 &   48  \\	
SC2\_18 	& 3:32:17.7 & 30:49:46 &    5.38 	&  2.95 	&  18.7 &  14.6 &   45  \\	
SC2\_19 	& 3:33:16.1 & 31:07:52 &    0.62 	&   0.42 	&  27.1 &  19.1 &   17  \\
SC2\_20$^{b}$   & 3:33:21.3 & 31:07:29 & 5.78 	&   2.23 	&  14.9 &  14.6 &   21  \\ 
SC2\_21  	&  3:43:51.2 & 32:03:22 &   1.18 	&   0.56 	&  25.3 &  14.6 &  180 \\	
SC2\_22   &  3:43:57.2 & 32:03:05 &   4.90 	&   1.32 	&  14.6 &  14.6 &   56  \\	
SC2\_23 	&  3:44:01.5 & 32:01:53 &   0.88 	&   0.82 	&  32.4 &  20.9 &   48  \\	
SC2\_24   &  3:44:44.1 & 32:01:34 &   2.32  	&   0.63 	&  14.6 &  14.6 &  129  \\	
\hline
\end{tabular}
%\end{center}
%\tablefoot{
\begin{tablenotes}[normal,flushleft]
\item \tnote{a} Peak 850 \um\ flux density ($S_{peak}$), total 850 \um\ flux density ($S_{total}$), semi-major axis ($a$), semi-minor axis ($b$), and position angle ($\theta$) from the \emph{getsources} extractions.  The core shape is not deconvolved with the beam. \tnote{b} The B1-b core was split into two objects with \emph{getsources}. We use the brighter core, centered with B1-bS (see text).
\end{tablenotes}
\end{threeparttable}
\end{table*}	

\begin{figure}
\includegraphics[width=\columnwidth]{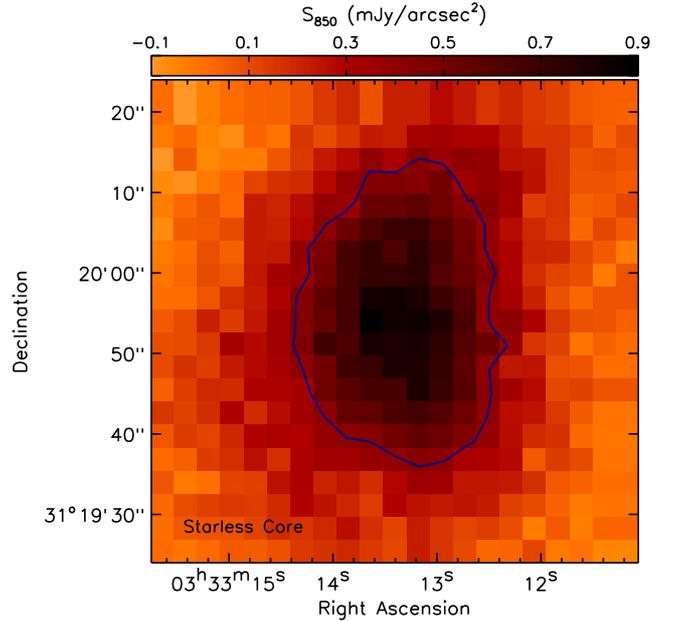}
\caption{An example starless core in Perseus at 850 \um.  The map is $\sim 1$\arcmin\ on a side, centered at the position of the core.  Blue contours show a flux density of 0.47 \mJyarcsec, which corresponds to 50\%\ of the peak core flux.\label{starless}}
\end{figure}

Roughly half of the binary systems are associated with elongated structures, judging from their 50\%\ flux contours (see Figure \ref{fig:allfeature1}).  Although such elongation could reflect a change in their density distribution due to the presence of embedded binaries, we similarly see elongated cores at the starless stage \citep[e.g., see Figure 13 in][]{Enoch06}.  Figure \ref{starless} shows a typically elongated starless core.  Although we have not conducted a systematic survey of starless cores in the Perseus Molecular Cloud, we speculate, based on such examples, that elongated cores which currently host binaries were already elongated prior to the onset of star formation \citep[e.g.,][]{Myers91}.   

\subsection{Coevality within Binaries}\label{noncoeval}

Most of the multiple systems in Perseus appear coeval, i.e., their embedded components have similar spectral energy distribution (SED) shapes.  Nevertheless, \citet{Murillo16} showed that several systems have components with different SED shapes and thus, may be non-coeval.  In our sample, the apparent non-coeval systems are SC2\_4, SC2\_11, SC2\_13, SC2\_16, SC2\_20, SC2\_23, and SC2\_24 (see Figure \ref{fig:allfeature1})\footnote{\citet{Murillo16} has an additional non-coeval system that involves Per-emb-10 and Per-emb-6 in NGC 1333.  These two objects are not considered binaries in \citet{Tobin16}, and similarly we find them associated with unrelated cores in our SCUBA-2 data.  Therefore, we consider these two objects to be two individual systems.}.

In several cases, we identify the ``non-coeval'' components as physically separate systems, associated with distinct cores (see Section \ref{coreID}).   Since most of the non-coeval systems are in NGC 1333 and IC 348, where the source number densities are higher \citep{Bally08, Gutermuth09}, such chance coincidences are more likely. The existence of these ostensibly non-coeval systems highlights the problem with using fixed separations to identify binary systems, particularly in clustered environments.  In contrast, we remove many of these false associations by identifying multiples based on the association of the stars with a single dense core.  

Even following our procedure, there remain five cores (SC2\_4, SC2\_5, SC2\_11, SC2\_16, and SC2\_24) that appear to contain non-coeval stars \citep[see Table \ref{coreBinary},][]{Murillo16}.  Most of these systems (SC2\_4, SC2\_5, SC2\_16, and SC2\_24) have widely separated components, where the stars nearer the core center are Class 0 and the outer stars near the 50\%\ flux contour are Class I or Class II.  The outer stars in these systems could be genuinely older, or simply appear older because they are observed through less extinction.  We assume the latter is true, i.e., that these four systems are each coeval and Class 0 based on the classification of their centermost star.    The remaining system, SC2\_11, contains four stars above the 80\%\ flux contour, so they should have similar extinction corrections.   Nevertheless, its SVS13A component is saturated in \emph{Spitzer} observations, and also blended with the other three components, making it difficult to disentangle their SEDs.   The global SED of this system is manifestly Class I \citep{Evans09,Sadavoy14}, so we take all four components to be at this phase.  

As previously noted, such anomalous systems could also arise if one star in the pair were born first and then drifted toward the cloud edge, after which its partner was born close to the core center. \citet{Foster15} found that Class II stars in NGC 1333 have higher intrinsic velocities than the dense cores in this region. However, these stars could have been free of embedding gas for several Myr, giving them ample time to acquire their larger, observed speeds.  We deem it unlikely that one star within a binary pair was born with such high speeds. Thus, we continue to assume that the mixed systems in our sample are truly coeval, with an age given by the more central component.

\section{Relation of Binaries to Dense Cores}\label{results}

\subsection{Spatial Location}

\begin{figure*}
\includegraphics[width=0.78\textwidth]{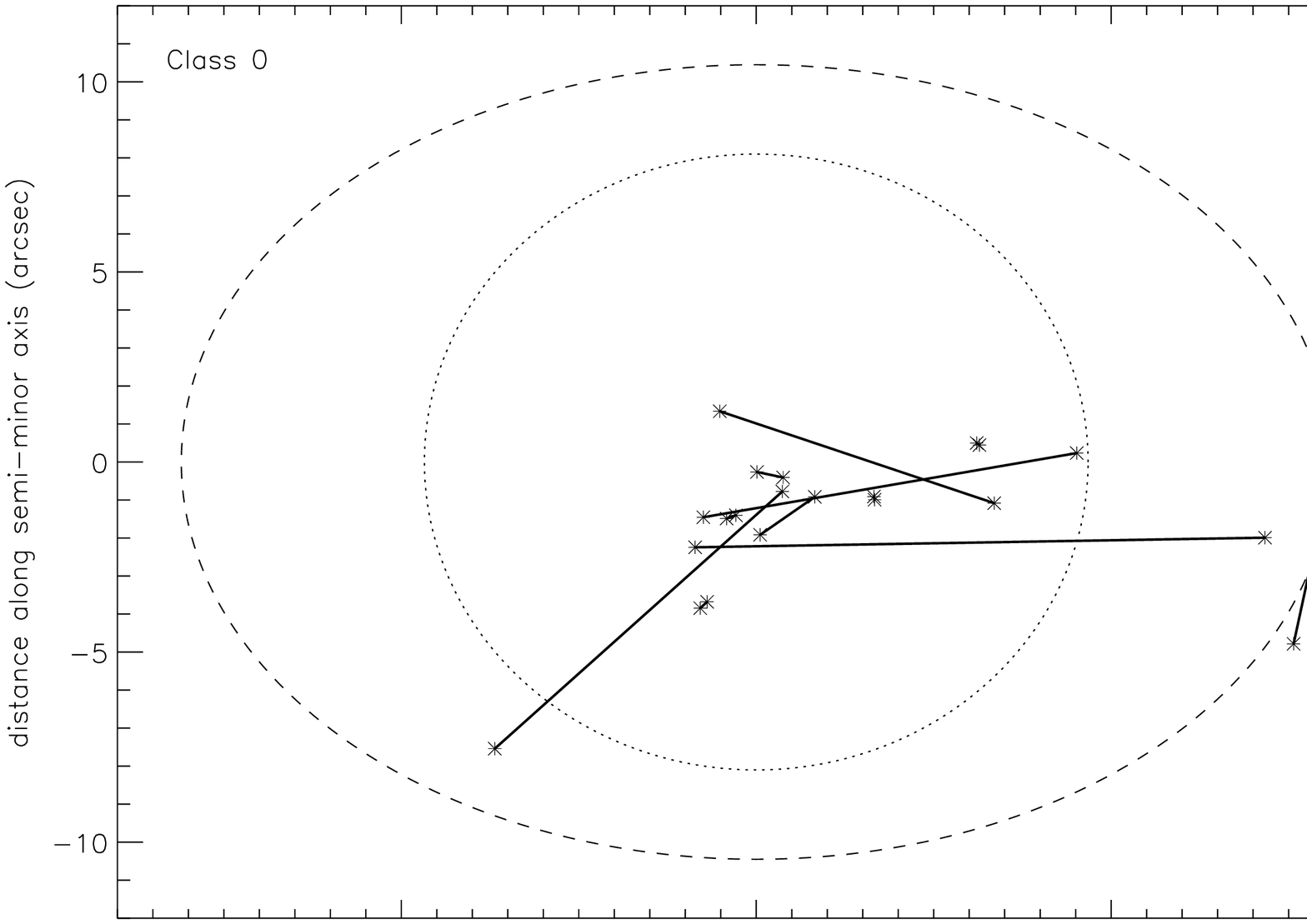}\\
\includegraphics[width=0.78\textwidth]{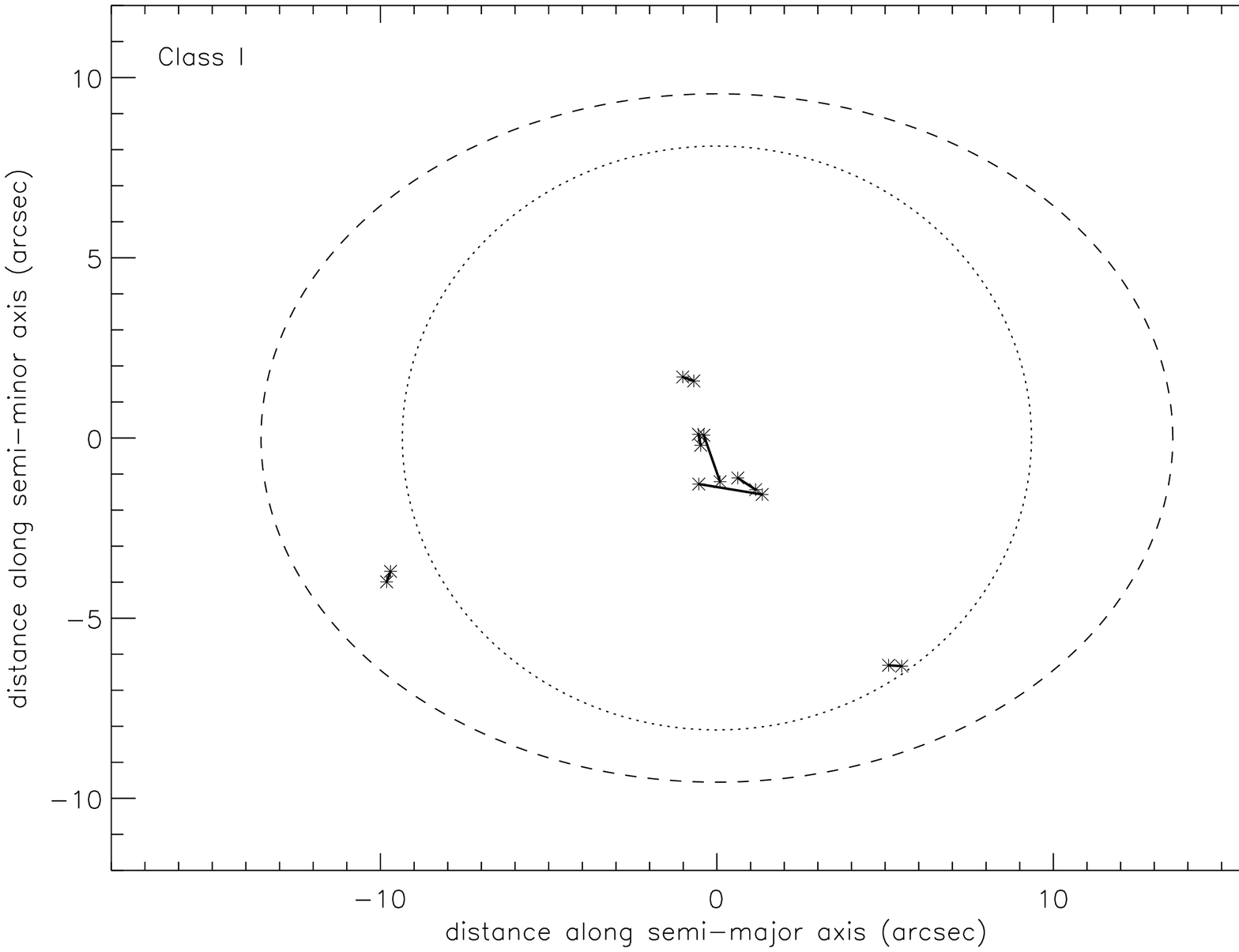}
\caption{Locations of embedded binaries within their parent dense cores for Class 0 systems (top) and Class I systems (bottom).  Stars show the position of the binary components along the semi-major and semi-minor axes of the host core with lines connecting them.  The dotted ellipses represent the mean core size for the Class 0 and Class I systems combined.  The dashed ellipses show the largest core size for Class 0 and Class I systems, respectively.  The core sizes have not been deconvolved with the beam (14.6\arcsec). \label{positions}}
\end{figure*}

\citet{Jorgensen07} showed that most embedded sources in Perseus are located within the central 15\arcsec\ of their cores.  Similarly, we find that most of our embedded systems are near the centers of their parent cores.  Figure \ref{positions} shows the positions of the embedded (2-star) binaries identified as Class 0 binaries (top) and Class I binaries (bottom), with lines connecting the binary components.  The binary positions are plotted relative to the semi-major and semi-minor axes of their parent cores to illustrate how each system is located within its own core and also relative to all other systems.  The dotted ellipses represent the mean core shape for all the embedded binaries (Class 0 and Class I) and the dashed ellipses represent the largest cores containing Class 0 and Class I binaries, respectively.  We see that the Class 0 systems include both wide and tight binary binaries, whereas the Class I binaries have only tight separations \citep[see also][]{Tobin16}.   

\begin{figure*}
\includegraphics[width=0.78\textwidth]{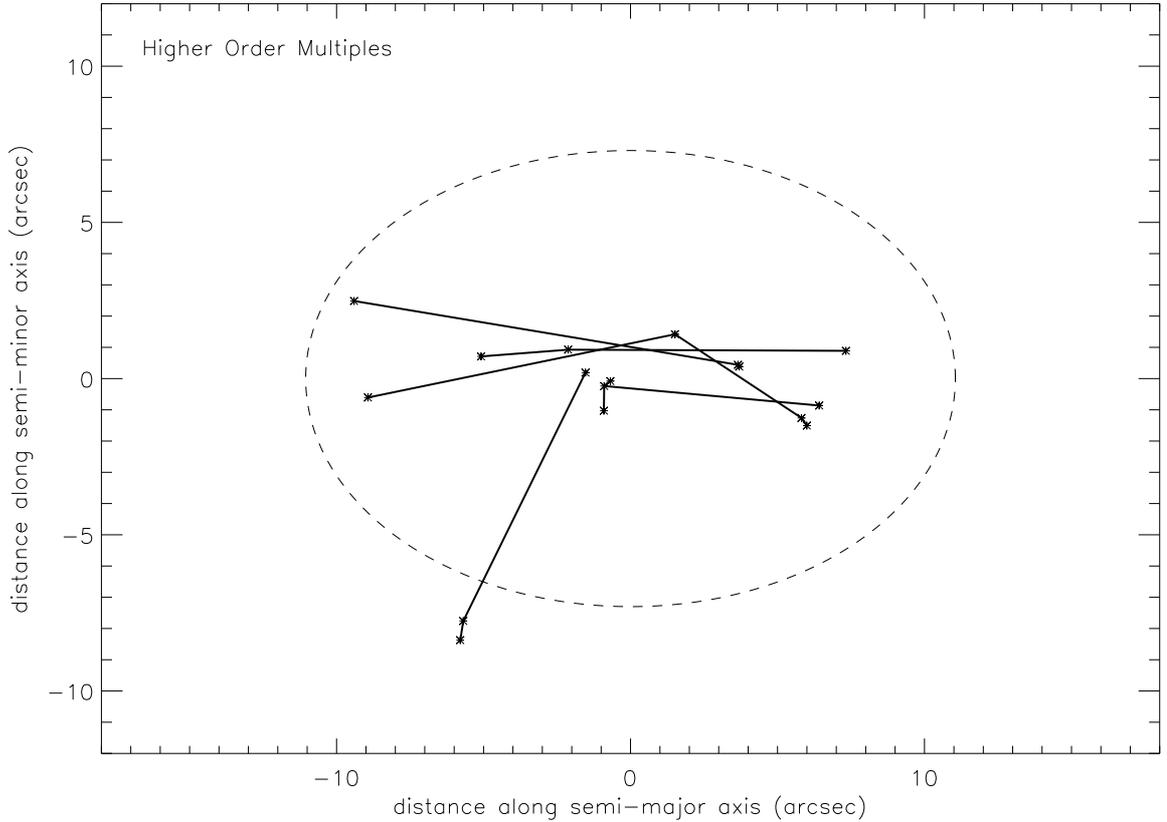}
\caption{Same as Figure \ref{positions}, but for the five higher-order multiples in Perseus.  Four of these systems are at the Class 0 phase and one is at the Class I phase.  The dashed ellipses show the largest core size for the higher-order multiples.  This core size has not been deconvolved with the beam (14.6\arcsec). \label{positionsHigh}}
\end{figure*}

Our sample also includes five higher-order multiples, i.e., systems with three or more components.  Four of these higher-order multiples are Class 0 systems, whereas only one (quadruple) system, SC2\_11, is at the Class I phase and is the only Class I system with widely separated stars.  Figure \ref{positionsHigh} shows the relative positions of the higher-order multiple components in the same manner as Figure \ref{positions}.  The higher-order multiples generally contain a mix of components at wide and tight separations, and all systems have at least one component far from the center of the core (see also Figure \ref{fig:allfeature1}).  It is interesting that the higher-order multiples are usually nested inside smaller cores than either the Class 0 or Class I binaries.  Only one higher-order multiple core (SC2\_13) is larger than the mean core for all embedded systems, and this core is still 40-50\%\ smaller than the largest cores for the Class 0 and Class I binaries.  Thus, larger stellar systems do not necessarily form from larger cores.

\subsection{System Orientation}\label{orientation}

Figures \ref{positions} and \ref{positionsHigh} show that most of the Class 0 binaries have their components along the semi-major axes of their parent cores.  Only two binaries (SC2\_16, SC2\_23) deviate from this trend, and in both cases, one or more components are located toward the edge of the core (see Figure \ref{fig:allfeature1}).  Thus, deeply embedded binaries appear to show a preferential alignment with the long axes of their host cores.  

\begin{figure}
\includegraphics[width=\columnwidth]{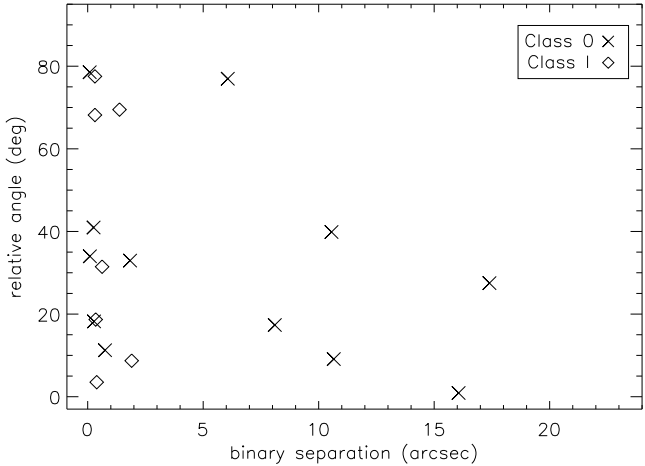}
\caption{Distribution of separations between pairs of binary stars with the relative angle between that binary pair and the semi-major axis of the parent core.  Class 0 systems are shown as crosses and Class I systems are shown as diamonds.  \label{angleVSsep}}
\end{figure}

To quantify this alignment, we measure the relative angle between each binary pair and the long axis of its parent core; relative angles of 0\degree\ correspond to systems aligned with the long axis and angles of 90\degree\ correspond to systems perpendicular to the long axis.  Figure \ref{angleVSsep} compares these relative angles with the star separation for the Class 0 and Class I binaries only.  

We see two distinct populations in Figure \ref{angleVSsep}.  At separations less than 2\arcsec\ ($\sim$ 500 au), the binaries appear to have no preferred angle, whereas at greater separations, the binaries are mostly aligned with their cores' long axes.    Of the six relatively wide Class 0 binaries, five of them have orientation angles less than 45$\degree$ and four of them have orientation angles less than 30$\degree$.  The two systems with angles greater than $30$\degree\ (SC2\_16, SC2\_23) have one or both stellar components near the edge of the core (see Figure \ref{positions} and Appendix \ref{appendix}).

\begin{figure}
\includegraphics[width=\columnwidth]{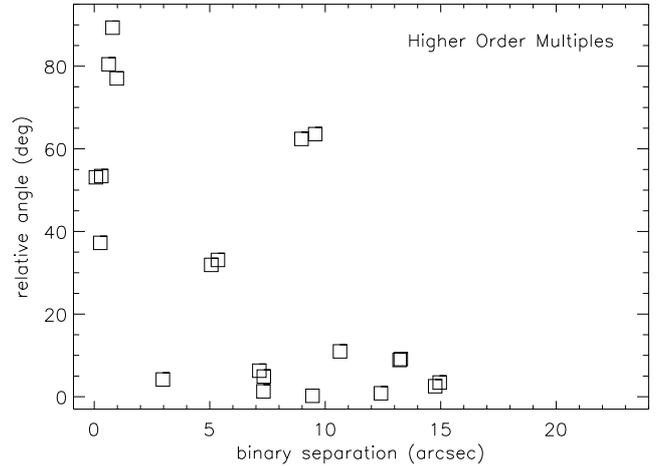}
\caption{Same as Figure \ref{angleVSsep} but for all unique stellar pairs within the higher-order multiple systems. \label{angleVSsepHigh}}
\end{figure}

We see a similar trend in the higher-order multiples.  Within each system, we considered all unique stellar pairs, and again plotted their relative angles as a function of separation.  (For example, a triple system has three such pairs.)  Figure \ref{angleVSsepHigh} shows the result.  As with true (two-star) binaries, the stellar pairs in the higher-order multiples exceeding $\sim 500$ au are generally aligned with the long axes of their host cores.  Tighter pairs are not so aligned, but rather have a tendency to be perpendicular to their core axes.    

The observed orientations of binary pairs relative to the long axes of their host cores are, of course, 2-dimensional projections of 3-dimensional systems on the plane of the sky. To connect these observations with true, 3-dimensional orientations, we performed Monte Carlo simulations of 1000 binary pairs with random positions within a generic, ellipsoidal dense core. We found that the binary pairs that were nearly parallel to the core axis (0-20$\degree$) remained so in projection.   In contrast, systems that were more orthogonal to the long axis had much broader variation in their projected orientations.

Thus, the trends seen in Figures \ref{angleVSsep} and \ref{angleVSsepHigh} are best reproduced if the wide binaries are aligned with the long axis of their host core and the tight binaries are either randomly aligned with or perpendicular to the long axis of their host core.  While all our conclusions must naturally be viewed with caution because of our small sample size, our observations strongly hint at a distinct difference in binary orientations when the stellar separation reaches about 500 au. 

\section{Theoretical Interpretation}\label{theory}

\subsection{Phenomenological Models}\label{models}

\subsubsection{Constructing the Models}

Several previous studies have found two distinct evolutionary trends in the populations of embedded multiple sources.  First, the overall fraction of multiple systems decreases with age, i.e., from Class 0 to Class I \citep{Looney00, Chen13, Tobin16}.  Second, there are more relatively wide binaries in the Class 0 phase than in the Class I phase (compare again Figures 3 and 4).   These findings, taken together, suggest that embedded binaries, especially those that are initially at wider separations, somehow become disrupted.  In addition, the prevalence of tighter binaries in the Class I phase may be telling us that systems which do \emph{not} breakup gradually shrink with time.

We now explore, in a simple but quantitative manner, how well the processes of binary orbital decay and disruption account for the observations in Perseus.  The observed data to be explained include the numbers and evolutionary stages of both binary and single sources.  We exclude the relatively few higher-order multiples, whose evolution we do not attempt to model.  As before, the number counts for both single and binary sources come from \citet{Tobin16}.  We also adopt the evolutionary stage assigned by these authors, adjusting the mixed systems (those containing both Class 0 and Class I sources) as previously described in Section \ref{noncoeval}.  Finally, we distinguish tight binaries from wide binaries using our threshold separation of 500 au from Section \ref{orientation}. 

Table \ref{resultsTable} gives the populations of wide, tight, and individual Class 0 and Class I systems in Perseus.  The first row indicates the observed numbers of each type.  In our notation, $N_0$ denotes the number of single, Class 0 sources, $N_{0,w}$ the wide, Class 0 binaries, and $N_{0,t}$ the tight, Class 0 binaries.  We use analogous notations for the Class I sources.  For all six populations, we also indicate our estimated errors in the number counts.  We derive these errors by considering both alternative evolutionary classifications from the literature \citep[e.g.,][]{SadavoyPhD, Sadavoy14, Murillo16}, and by provisionally adopting a tight-wide threshold of 300 au, which equally fits the separation distributions shown in \citet{Tobin16}. 

\begin{table}
\begin{threeparttable}
\caption{Observed Population Counts and Model Values}\label{resultsTable}
\begin{tabular}{lcccccc}
\hline\hline
Population			& $N_0$		&$N_{0,w}$	& $N_{0,t}$	& \NI		& \NIw		& \NIt \\
\hline\\[-3mm]
Observed				& $15^{+1}_{-1}$	& $6^{+1}_{-3}$& $6^{+1}_{-2}$&  $30^{+1}_{-4}$ & $0^{+4}_{\mbox{---}}$	& $7^{+1}_{-3}$  \\[1mm]
%\hline
%wi\_s model			& 14.1		& 7.5			& 6.5			& 12.0	& 3.7			& 8.3 \\		
%wti\_b model			& 15.0		& 1.6			& 6.1			& 23.7	& 0.5			& 8.2 \\
w\_sb model			& 15.1		& 5.3			& 3.8			& 28.8	& 1.9			& 7.3 \\
\hline
\end{tabular}
\begin{tablenotes}[normal,flushleft]
\item Observed and modeled source counts exclude higher-order multiples.  See text for details.
\end{tablenotes}
\end{threeparttable}
\end{table}

We next construct a series of phenomenological models that include varying degrees of binary shrinking and breakup.  We defer our assessment of the physical basis for both processes to Section \ref{basis}.  In all our models, we assume that all sources form in the Class 0 phase as either single stars or binaries.  We denote their formation rate (number per time) as $R$, measured in Myr$^{-1}$.  All stars then evolve to the Class I phase over a characteristic lifetime, \tOI.  During this evolution,  \emph{wide} binaries in either the Class 0 or Class I phase can shrink into tight binaries over a shrinking time scale, $t_s$, or else break up into two individual stars over a breakup time scale, $t_b$. Finally, the Class I stars themselves evolve to the Class II phase in a characteristic lifetime, \tI.  For simplicity, we assume that $R$, $t_s$, $t_b$, \tOI, and \tI\ are themselves constant in time.

When a binary breaks up into two single stars, the latter cannot both remain in the same dense core, or the system would still be considered a wide binary observationally.  Thus, in our models that include binary breakup, we are assuming that this process occurs when the dense core itself splits into two, with each portion containing a very young star. We will consider further the astrophysical basis of this assumption in Section \ref{break} below.  Alternatively, the dense core may remain intact, but one or both of the stars may somehow be ejected.  In such a scenario, the ejected stars would be observed in a more advanced evolutionary phase than Class 0 or I.  This alternative picture can be described through ``ejection models'', which we detail and evaluate in Appendix \ref{appendixA}.  We assess the physical bases of these two pictures of breakup in Section \ref{break}. 

Table \ref{modelTable} lists the models we tested for the binary breakup scenario.  These represent all the ways by which the cloud can create individual stars and binaries, which then evolve to the observed populations.  The first column identifies the models, using a convenient shorthand notation of the form x\_y.  Here, x represents the kinds of systems formed in the cloud.  These can be wide binaries (w), tight binaries (t), or individual stars (i).  The symbol y gives the processes to which the wide binaries are subject, either shrinking (s) or breakup (b).  Thus, the first model listed in the table, denoted wi\_s, is one in which both wide binaries and individual stars form in the cloud at the rate $R$, and the wide binaries can only shrink with time.  For clarification, the next five columns of the table give explicitly the types of systems formed and the fate of the wide binaries for each model.  For the stellar ejection scenario (see Appendix \ref{appendixA}), we add star symbols to indicate the number of stars ejected from the core during breakup.

\begin{table}
\begin{threeparttable}
\caption{Comparison of Different Models}\label{modelTable}
\begin{tabular}{lcccccc}%{lcccccccc}
\hline\hline
\multirow{2}{*}{Model} 	&  \multicolumn{3}{c}{Formed in the Cloud}	&  \multirow{2}{*}{shrink} & \multirow{2}{*}{break} & \multirow{2}{*}{$\chi^2$}  \\								%shrink			breakup
		&  wide	     	   & tight	         	& indiv.	      	   &		       		&	 			& \\
\hline
wi\_s		& $\checkmark$  & 		         	& $\checkmark$   & $\checkmark$  	& 				& 26 \\ %done
wti\_s	& $\checkmark$  & $\checkmark$     & $\checkmark$   & $\checkmark$  	& 				& 89 \\ %done
wt\_b	& $\checkmark$  &$\checkmark$  	&			   &		       		& $\checkmark$  	& 19\\ %done
wti\_b	& $\checkmark$  & $\checkmark$  	& $\checkmark$  &		        		& $\checkmark$  	& 6.1 \\ %done
w\_sb	& $\checkmark$  &		         	&			   & $\checkmark$  	& $\checkmark$  	& 1.6 \\ %done
wt\_sb	& $\checkmark$  & $\checkmark$  	&			   & $\checkmark$  	& $\checkmark$  	& 19 \\ %done
wi\_sb 	& $\checkmark$  & 		         	& $\checkmark$  & $\checkmark$  	& $\checkmark$ 	& 3.1 \\ %done
wti\_sb	& $\checkmark$  & $\checkmark$  	& $\checkmark$  & $\checkmark$  	& $\checkmark$  	& 6.1 \\ %done
\hline
\end{tabular}
\end{threeparttable}
\end{table}

\subsubsection{Solving the Models}\label{sfr}

To evaluate each model quantitatively, we set up a series of linear differential equations describing the evolution of the various populations. Consider, for example, the model w\_sb. (This will turn out to be the optimal one; see below.) In this model, the cloud produces only wide binaries. After the binaries form, they both shrink and break apart, at relative rates determined by the respective time scales. Thus, the governing equations for this particular model are:
\begin{subequations} \label{sb_main}
\begin{eqnarray}
dN_{0,w}/dt &=& R - N_{0,w}/t_s - N_{0,w}/\tOI - N_{0,w}/t_b   \\
dN_{0,t}/dt &=& N_{0,w}/t_s - N_{0,t}/\tOI   \\
d\NIw/dt &=& N_{0,w}/\tOI - \NIw/t_s  - \NIw/t_b - \\
	      & &	\NIw/\tI \nonumber   \\
d\NIt/dt &=& \NIw/t_s + N_{0,t}/\tOI - \NIt/\tI  \\
dN_{0}/dt &=& 2N_{0,w}/t_b - N_{0}/\tOI   \\
d\NI/dt &=& 2\NIw/t_b + N_{0}/\tOI - \NI/\tI
\end{eqnarray}
\end{subequations}

\noindent All our equations govern only the embedded (Class 0 and I) stars.  We do not track the evolution of more advanced phases, because their multiple star populations are incomplete\footnote{For example, \emph{Spitzer} surveys of Perseus have resolutions greater than $2$\arcsec\ (470 au) and will be unable to resolve most tight binaries \citep[e.g.,][]{Cieza09}.}.

The above equations contain five free parameters: $R$, $t_s$, $t_b$, \tOI, and \tI.  We can reduce the parameter space by dividing all equations by the time \tOI.  Defining a non-dimensional time by $\tau = t/\tOI$, we find a new set of equations:
\begin{subequations} \label{sb_dimension}
\begin{eqnarray}
dN_{0,w}/d\tau &=& (R\ \tOI) - N_{0,w}/(t_s/\tOI) - N_{0,w} - \\
			& &  N_{0,w}/(t_b/\tOI)  \nonumber \\
dN_{0,t}/d\tau &=& N_{0,w}/(t_s/\tOI) - N_{0,t}   \\
d\NIw/d\tau &=& N_{0,w} - \NIw/(t_s/\tOI)  - \\
			& & \NIw/(t_b/\tOI) - \NIw/(\tI/\tOI) \nonumber  \\
d\NIt/d\tau &=& \NIw/(t_s/\tOI) + N_{0,t}  - \NIt/(\tI/\tOI)  \hspace{2mm}\\
dN_{0}/d\tau &=& 2N_{0,w}/(t_b/\tOI) - N_{0}   \\
d\NI/d\tau &=& 2\NIw/(t_b/\tOI) + N_{0} - \NI/(\tI/\tOI)
\end{eqnarray}
\end{subequations}

\noindent Solving this system requires choosing values for only four non-dimensional quantities: ($R\ \tOI$), $t_s/\tOI$, $t_b/\tOI$, and \tI/\tOI.  In practice, the embedded lifetimes are brief (under 0.5 Myr) compared to the maximum age of observed stars in Perseus \citep[2 to 3 Myr;][]{Bally08}.  Note that Perseus contains over 240 Class II sources, which is significantly more than the 76 embedded systems \citep{Dunham15}.   Therefore, the latter should have attained steady-state populations, representing a balance between their creation and destruction, as effected by their evolution to more advanced stages or through binary breakup and shrinking. 

We apply this steady-state assumption by setting all the time derivatives in the appropriate evolutionary equations to zero.  We thus obtain six algebraic expressions for the various populations in terms of the four unknown non-dimensional parameters.  For each set of parameter values, we compare the resulting steady-state populations with the observed ones given in Table \ref{resultsTable}, and compute an error $\chi^2$.  We then vary the parameters to obtain the lowest $\chi^2$ for all eight models.  These $\chi^2$ values are listed in the rightmost column of Table \ref{modelTable}.   Note that the best-fit versions of the wt\_sb and wti\_sb models both have very long times for orbital shrinking, which makes these models identical in practice to the wt\_b and wti\_b cases, respectively.  The w\_sb model yields the lowest error, as claimed previously.   Again, this model posits that the cloud produces only wide, Class 0 binaries which then either shrink and break apart as they evolve to the Class I phase.

The non-dimensional parameter values in the w\_sb model are $R\ \tOI = 16.74$, $t_s/\tOI\ = 1.39$, $t_b/\tOI\ = 0.70$, and $\tI/\tOI = 1.41$.  Note that the time scale for binary shrinking exceeds the binary breakup time by about a factor of two.  Finding a dimensional rate $R$ and dimensional time scales for $t_s$, $t_b$, and \tI\ require that we select a value for \tOI\  based on observations.  \citet{Dunham15} estimated the Class 0 lifetime to be $0.13-0.26$ Myr, using number counts of stellar populations and an assumed pre-main-sequence lifetime of $2-3$ Myr.  We adopt the mean value obtained by this study, $\tOI = 0.2$ Myr. 

\begin{figure}
\includegraphics[width=\columnwidth]{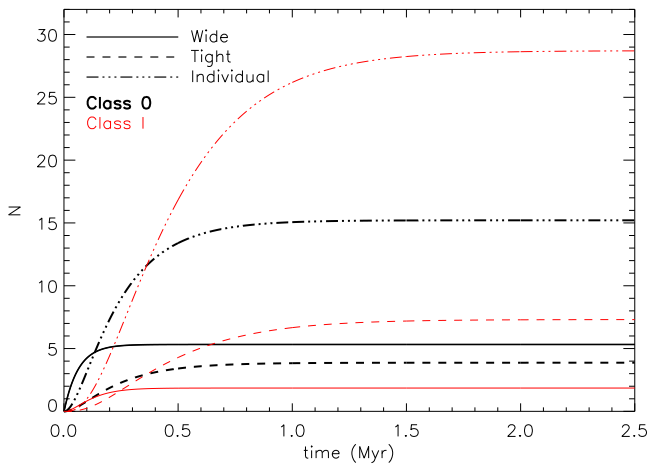}
\caption{Evolution of binary and single systems for our w\_sb model (see Equation \ref{sb_dimension} and the text) for $\tOI = 0.2$ Myr.  Wide binaries are shown by solid lines, tight binaries are dashed lines, and individual sources are dotted lines with Class 0 systems in (thick) black and Class I systems in (thin) red.    \label{model_pops}}
\end{figure}

Supplied with our chosen value of $\tOI$, we may now numerically integrate Equations (\ref{sb_dimension}a) through (\ref{sb_dimension}f) to obtain the temporal dependence of all populations within the w\_sb model.  Figure \ref{model_pops} displays this result.  We see that all  embedded populations indeed reach a steady state, by about 1.5 Myr.  The numbers of Class 0 objects, both binaries and individual stars, level off relatively quickly, within about 0.5 Myr.   In contrast, the Class I tight binary and single star populations take more time to asymptote.   All steady-state populations in this model agree well with observations (see the second line in Table \ref{resultsTable}). 

In solving Equations (\ref{sb_dimension}a)-(\ref{sb_dimension}f), we have also determined the total rate at which the cloud produces stars.  For the optimal w\_sb model, the production rate of wide binaries is $R = 84$ Myr$^{-1}$, a value that is again based on an assumed Class 0 lifetime of 0.2 Myr.  Since each binary produces two stars, we find a stellar production rate of 168 Myr$^{-1}$.  Previously, \citet{Evans09} derived a formation rate in Perseus by counting the total number of young stars (embedded and visible) and dividing by the duration of star formation, taken to be 2 Myr.  They used \emph{Spitzer} observations to identify 385 sources, and derived a similar stellar production rate of 193 Myr$^{-1}$ \citep[see also][]{Hatchell07, HsiehLai13, Young15}.  We find this agreement encouraging, but caution that it may be somewhat fortuitous.  Specifically, both our analyses may underestimate the true production rate, although not to a great degree.  \citet{Evans09} recognized that an unknown fraction of their sources were binaries unresolved by \emph{Spitzer}, and that counting all the individual components would boost their birth rates.  (Many of these tighter binaries were subsequently resolved by the VANDAM survey on which we have based our study.)  For our part, the calculated model excludes higher-order multiples for simplicity.  We do not attempt any adjustments of the stellar production rates, but simply note their consistency.

\subsection{Physical Basis}\label{basis}

In our simple numerical models, we formed binaries and allowed the wide systems to shrink into tight systems or break up into individual stars without giving any physical basis for these processes.  Here we discuss the physical processes for binary formation and dynamical evolution, and connect these predictions to our observations.

\subsubsection{Binary Formation}\label{formation}

Within our set of phenomenological models, the one that best matches the Perseus observations posits that all stars initially form as wide binaries. Assuming this conclusion will still hold after studying other regions, we now briefly examine current theories of binary formation and assess their compatibility with our observations. 

Two leading ideas are turbulent and rotational fragmentation of the parent dense core. Both models have been explored through numerical collapse simulations. In the turbulent picture \citep[e.g.,][]{Goodwin04, Offner10, Offner16}, the dense core is taken to be a spherical structure endowed with varying degrees of turbulence.  Self-gravity in this initial state is stronger than support from internal pressure, and the cloud immediately collapses to a flattened configuration, which then breaks apart into a number of relatively dense fragments.  In the rotational picture \citep[e.g.,][]{Hennebelle04, Machida08, Klapp14}, the initial cloud is again a strongly self-gravitating sphere, but now given a modest degree of rotation.  As before, the sphere flattens as it collapses, and subsequently fragments once it attains a ring- or bar-like structure.  

The foregoing simulations all assume that dense cores are perfect spheres, but deviations from this idealized geometry may influence collapse and fragmentation.  Since we observe only the two-dimensional projections of dense cores, their intrinsic shapes must be inferred through statistical analysis. Restricting themselves to axisymmetric configurations, \citet{Myers91} concluded that dense cores are prolate ellipsoids \citep[see also][]{Ryden96, Curry02}.   \citet{JonesBasu02} relaxed the assumption of axisymmetry, and claimed that triaxial shapes give a better match to observations.  However, the more recent finding that many cores lie along chains within larger filaments provides additional support for prolate configurations \citep[e.g.,][]{TafallaHacar15}.  To simulate non-spherical collapse, \citet{Bonnell91} modeled dense cores as cylinders.  These cylindrical cores collapse transversely to their main axes, creating narrow spindles that subsequently fragment and fall toward one another \citep[see also][]{Pon11}.

Other researchers have explored the production of binaries from a single star and its disk, where fragmentation of the latter creates the secondary component.   Within this picture, \citet{StamatellosWhitworth09} followed the breakup of a disk with mass equal to that of the central star.  \citet{Kratter10} performed a suite of simulations, examining the tendency for fragmentation in young, rapidly accreting disks being built up at varying rates from the parent cloud. 

The simulations we have cited span many years and a broad range of initial conditions.  Nevertheless, they all share one key prediction: binaries arise from relatively small substructures (e.g., rings, bars, disks, or spindles) created by global cloud collapse of the original  dense core.  The observations do not support this view, in that we find no evidence for such features in the dense cores hosting widely separated, Class 0 binaries.  However, with the modest angular resolution of SCUBA-2 ($\sim$ 3400 au at the distance of Perseus), we could be missing relatively compact and denser structures containing the stars.  Future observations with both greater sensitivity and resolution will be valuable in revealing the spatial distribution of material in the neighbourhood of the embedded stars.

In this context, we note the interesting, recent study by \citet{Tobin16l1448} of the dense core SC2\_4.  This core hosts a hierarchical quadruple, of which three Class 0 stars have separations of order 100 au (see Figure \ref{fig:allfeature1}).  \citet{Tobin16l1448} used ALMA to image these three objects, Per-emb-33-A, -B, and -C, in the dust continuum at 1.3 mm.  They found that stars A and B are driving a pair of spiral waves in a circumbinary disk, while star C is located in the periphery of this structure.  After analyzing the gravitational stability of the disk, Tobin et al. argue that it recently created star C through fragmentation.  Alternatively, all three stars could have been born with much larger mutual separation, and then drifted together. Such a picture is more consistent with our model for binaries, but is admittedly more speculative in the case of higher-order multiples, which the models exclude.

\subsubsection{Shrinking Orbits}\label{shrink}

The components of an embedded binary interact gravitationally both with each other and with their host dense core.  Their motion will depend on which force dominates. The top panel of Figure \ref{positions} shows that some Class 0 binaries still have orbital separations comparable to the core diameter. As one might expect, these stars predominantly feel the gravitational force from the core. We may bolster this conclusion through the following, simplified analysis.

We consider the forces acting on a binary situated inside a dense core,  where the latter is idealized to be spherical. Let $r_0$ be the core radius and $\rho$ the mean density. Then the total core mass is
\begin{equation}\label{eq1_Mcore}
M_{core} = (4\pi/3)\rho\ r_0^3.
\end{equation}
In the middle of this core are two identical, Class 0 stars of mass $M_{star}$, located on opposite sides of the core center, and each a distance $r_{star}$ from this point. We wish to find that critical value of $r_{star}$ such that the gravitational force between one star and its companion equals that between the star and the host core.  The first force is $F_1 = G M_{star}^2 / (4 r_{star}^2)$.  The second force arises only from that portion of the core mass internal to the star. Assuming the core has uniform density, this internal mass is
\begin{equation}
M_{int} = (4\pi/3)\rho\ r_{star}^3.
\end{equation}
The force of each star toward the entire core is then $F_2 = G M_{int} M_{star} / r_{star}^2$. After equating $F_1$ and $F_2$, we find that
\begin{equation}
M_{star}/4 = (4\pi/3)\rho\ r_{star}^3.
\end{equation}

To determine the critical value of $r_{star}$ when the forces balance, we must now relate the mass of the star to that of its host core.  Assessing Class 0 star masses is nontrivial, since the stars are deeply embedded in dusty gas.  Nevertheless, the stellar masses of a few sources have been obtained by analysis of their Keplerian disks \citep{Tobin12,  Murillo13, Codella14, Yen16}.  The results range from $0.2 - 0.3$ \Msun, compared to host core masses of $0.5 - 2$ \Msun\ \citep[see aforementioned papers and][]{Miettinen10}.  In contrast, Class I stars tend to have greater masses than their cores \citep[][and references therein]{Chou14}.  Based on these observations, we adopt the relation $M_{star} = M_{core} / 2$, to ensure that the total mass of stars in our Class 0 binaries does not exceed that of their host cores. We may then rewrite equation (5) as
\begin{equation}\label{eq2_Mcore}
M_{core}/8 = (4\pi/3)\rho\ r_{star}^3.
\end{equation}
After dividing Equation \ref{eq2_Mcore} by Equation \ref{eq1_Mcore}, we find
\begin{equation}
r_{star}/r_0 = 8^{-1/3} = 0.5.
\end{equation}

The numerical value for this ratio is actually an upper bound.   First, Class 0 stars may have masses that are \emph{less} than $M_{core}/2$.  Second, we have ignored the fact that the density distribution in the core is centrally peaked \citep[see, e.g.,][]{Jorgensen07}. Both corrections act to lower the ratio, but not by a lot. For example, if $M_{star}/M_{core}$ is correct, but the interior density is 10 times the mean (an extreme assumption), then $r_{star}/r_0$ falls to just 0.23.

The lesson, in any case, is that the widest Class 0 binaries, which are presumably the youngest, are bound to their host core and not to each other.  Nevertheless, the net force pulling on each star is inward\footnote{As we have noted, a more realistic representation of the dense core would be a prolate ellipsoid. The force on a star inside such a body is still only from the internal mass, as long as the isodensity contours are similar ellipsoids \citep{Chandrasekhar69}.} toward the companion star, causing the two stars to drift together even as they are gathering mass from the core. The stars initially share the rotational motion of the core, but their transverse speed subsequently increases by conservation of angular momentum.  Once their separation falls below the critical value, the stars attract each other and enter a true binary orbit.  Inward drift continues and accelerates as angular momentum is transferred to the background gas via dynamical friction.   See \citet{Stahler10} for an analytic treatment of this orbital decay.  

Simulations of orbital shrinking indicate the process primarily occurs at early times.  \citet{Offner16} explored binary formation and evolution in magnetohydrodynamical simulations, and found that binaries with separations of few 10$^3$ au could shrink to separations of 100 au in only 0.1 Myr.    Similarly, \citet{Bate02} found that wide binaries could produce very close binaries through accretion processes or via dynamical interactions while the stars are embedded in a gas-rich core. 

The optimal w\_sb model gives a shrinking time scale of $t_s/\tOI = 1.39$, which becomes $t_s \approx 0.28$ Myr for $\tOI = 0.2$ Myr.  (Note that the ejection models explored in Appendix \ref{appendixA} yield a similar shrinking rate for the optimal w\_sb$^{\ast}$ model.)  This shrinking rate is about three times longer than what was obtained by \citet{Offner16} for a 4 \Msun\ core.   Most of the cores in Perseus have masses of $\sim 1$ \Msun\ \citep{Sadavoy10}, so an average $t_s \approx 0.3$ Myr may still be in rough agreement with the simulations.    Both the w\_sb and w\_sb$^{\ast}$ models predict that binary shrinking occurs primarily during the embedded phase, as is also found by simulations.  Thus, our simple evolutionary models broadly agree with the theoretical studies currently available.  

Finally, our observations show that relatively tight binaries have random orientations with respect to their host cores, a situation that does not hold for the predominantly aligned wide binaries (recall Figure \ref{angleVSsep}). Thus, the processes that cause the wide binaries to shrink must also randomize their orbital planes.  Dynamical friction may also be responsible for this process.  The density inside a core is not perfectly smooth \citep[e.g.,][]{Looney07, Stutz09, Tobin10}, and this lumpiness will cause the torque from dynamical friction to change direction in time.  Whether this effect can truly yield the observed dispersion in orientation is unclear, but at least the process acts in the proper direction.

\subsubsection{Breakup}\label{break}

In our phenomenological model, newly formed binaries not only shrink with time, but also break apart into single stars.  Indeed, the time scale for breakup is generally faster than that for orbital shrinking, suggesting that breakup is a more efficient.  We considered two binary breakup scenarios, where the parent core itself splits into two parts (see Section \ref{models}) and where one or both of the stars are ejected from the system (see Appendix \ref{appendixA}).  We find two optimal models where the steady-state solution matches the observed populations of embedded stars: the w\_sb model ($\chi^2 = 1.6$) and the w\_sb$^{\ast}$ model ($\chi^2 = 1.5$).  The models where two stars are ejected do not fit the observations as robustly, and we exclude them from further discussion.  We are then left with two scenarios for binary breakup that equally reproduce the observations.    Therefore, we must use other considerations in judging the two.
 
The root cause of binary breakup is difficult to assess, and there is little guidance in the literature.  It has long been known that stellar dynamical interactions (e.g., within higher-order multiples) can lead to ejections from a bound system \citep[e.g.,][]{ReipurthClarke01, Reipurth14}.  In dense clusters, close passes of single or multiple stars can disrupt binaries and similarly lead to ejections \citep[e.g.,][]{Kroupa99, Parker09}.  These processes are less likely to occur when the gravitational potential is dominated by the dense core, as in our systems. 

Turning to the prospect of core breakup, we first note that cloud fragmentation lies at the heart of most current models for binary formation (recall Section \ref{formation}).  This process, however, is thought to occur before the stars themselves form.   Moreover, core fragmentation in these theories does not split the core into pieces, but creates substructures that later form stars.  Accordingly, we view global core breakup as unlikely.  It may be that stellar outflows erode so much gas that the parent core effective becomes two.  If this process occurs, then future observations should show a systematic difference in mass between cores containing embedded binaries and single stars.  At present, we lack the statistics to conduct this test.

\begin{figure}
\includegraphics[width=\columnwidth]{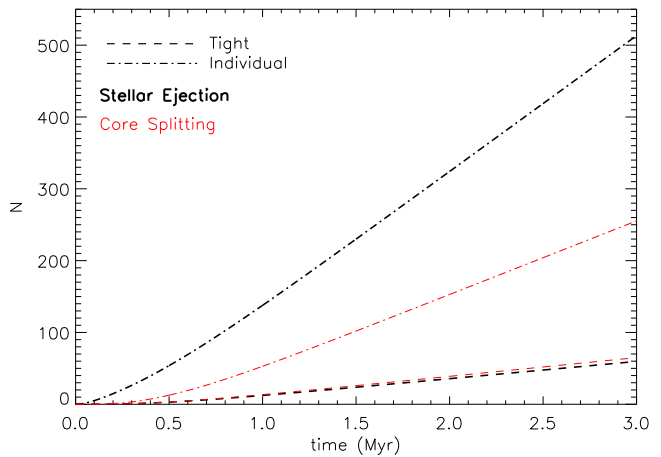}
\caption{Evolution of stars more evolved than Class I.  Tight binaries are shown by the two dashed lines (nearly overlapping) and the single stars are dot-dashed lines for the stellar ejection w\_sb$^{\ast}$ model in (thick) black and the core splitting w\_sb model in (thin) red.   As before, we assume $\tOI = 0.2$ Myr for all populations.   \label{evolved_fig}}
\end{figure}

Alternatively, the two binary breakup scenarios make very different predictions about the populations of more evolved stars.  Figure \ref{evolved_fig} shows the temporal dependence of single and tight evolved stars following the w\_sb and w\_sb$^{\ast}$ models.   Both models produce similar numbers of tight binaries (evolved primarily from tight Class I systems), but very different numbers for single stars.  The w\_sb$^{\ast}$ model produces more single, evolved stars than the w\_sb model because there is a high rate of stellar ejections ($t_b/\tOI = 0.37$).  Thus, further observations of the more evolved stars could favour one model over the other.

As mentioned previously, surveys with \emph{Spitzer} lacked the resolution to identify tight binary systems with separations $\lesssim 470$ au in Perseus \citep{Cieza09}.  Only a few subregions have higher resolution, near-infrared observations \citep[e.g., IC 348; ][]{Duchene99}.  We need deep, high-resolution observations of the evolved stars to ensure we have their complete population distributions.  In particular, stars ejected during the Class 0 phase may have very low masses.  Such stars may not be classified as more evolved sources, but their presence would support stellar ejection as the main mechanism behind binary breakup.

\section{Summary}\label{summary}

In this paper, we combine the uniform and complete binary database from the VANDAM survey \citep{Tobin16} with newly identified cores from SCUBA-2 observations at 850 \um\ \citep{MChen16} for the Perseus molecular cloud.  Our main results are:

\begin{enumerate}[label={(\arabic*)}, leftmargin=*]
\item Most of the embedded binaries are located toward the center of their parent cores.  Those few systems with one or more components along the edge of the core tend to have more tenuous (less compact) cores.\\[-2mm]
\item About half of the binaries are located within clearly elongated cores.  Since we find starless cores that are similarly elongated, we suggest that the embedded binaries inside elongated cores formed from dense, starless cores that were initially elongated.\\[-2mm]
\item Those binaries with separations greater than about 500 au tend to be aligned with the major axes of their parent cores.  In contrast, tighter binaries have more random orientations.   We find similar alignment in the relatively few higher-order multiples.\\[-2mm]
\item We have formulated a set of simple phenomenological models to describe the evolution of embedded stars, both single and binary.  Our best-fit model posits that all stars form as wide binaries, i.e., with separations exceeding 500 au. \\[-2mm]
\item Our model further predicts that most wide binaries break apart, although some shrink to become tighter systems.  We have considered two modes of binary breakup - core splitting and stellar ejection.  Under either mode, we obtain equally good fits to the observed populations of embedded stars.  Deeper and higher-resolution studies of more evolved stars may help decide which picture is correct.  \\[-2mm]
\item Within our best-fit models, we obtain a stellar birth rate of $168$ Myr$^{-1}$ for the core splitting scenario and 247 Myr$^{-1}$ for the stellar ejection scenario, assuming a Class 0 timescale of 0.2 Myr in both cases.  Either rate is reasonably close to that previously estimated by \citet{Evans09}.  \\[-2mm] 
\end{enumerate}

Our results predict that all stars are born initially as wide binaries.  This result agrees with the conjecture of \cite{Kroupa08} that most stars originate in binary systems.  Note, however, that this conjecture was based on the behaviour of binaries in N-body simulations lacking gas, whereas observations indicate that the first binaries arise in dense cores.  We therefore amend the conjecture to the following; \emph{Most stars form in wide binaries within dense cores.  These primordial binaries tend to be aligned with the long axes of their host cores.}

If most low-mass stars form initially as binaries, then we should expect each starless core to form two stars on average instead of the often assumed one-to-one correspondence.  Several studies have suggested star formation efficiencies of $\sim 30$\% based on a one-to-one correspondence between core mass functions (CMFs) and the initial mass function \citep[IMF; e.g., see ][]{Alves07}.  With each core forming two stars, fewer cores need to collapse to produce the current population of stars in clouds.  Instead, the mass fraction of dense cores that turns into stars could be double what is generally believed.

%%%%%%%%%%% ACKNOWLEDGEMENTS %%%%%%%%%%%%%
%\begin{center}
%\textbf{ACKNOWLEDGEMENTS}
%\end{center}
\vspace{1cm}
\section*{Acknowledgements}
The authors thank the anonymous referee.  SIS acknowledges the support for this work provided by NASA through Hubble Fellowship grant HST-HF2-51381.001-A awarded by the Space Telescope Science Institute, which is operated by the Association of Universities for Research in Astronomy, Inc., for NASA, under contract NAS 5-26555.  The authors thank the JCMT staff for their support of the GBS team in data collection and reduction efforts.  The authors acknowledge and the thank the following researchers for their contributions to the DR1 SCUBA-2 data used: J. Buckle, D. Berry, M. Currie, J. Hatchell, T. Jenness, D. Johnstone, H. Kirk, J. Mottram, K. M. Pattle, and S. Tisi.  The authors also thank T. Jenness, H. Kirk, M. Lombardi, S. Offner, and J. Tobin for comments and discussions.

The JCMT has historically been operated by the Joint Astronomy Centre on behalf of the Science and Technology Facilities Council of the United Kingdom, the National Research Council of Canada and the Netherlands Organization for Scientific Research. Additional funds for the construction of SCUBA-2 were provided by the Canada Foundation for Innovation.  The \emph{Starlink} software \citep{Currie14} is supported by the East Asian Observatory.  This research used the services of the Canadian Advanced Network for Astronomy Research (CANFAR) which in turn is supported by CANARIE, Compute Canada, University of Victoria, the National Research Council of Canada, and the Canadian Space Agency.  The authors wish to recognize and acknowledge the very significant cultural role and reverence that the summit of Maunakea has always had within the indigenous Hawaiian community. We are most fortunate to have the opportunity to conduct observations from this mountain. 

%%%%%%%%%%%   BIBLIOGRAPHY  %%%%%%%%%%%%%%%%
\bibliographystyle{mnras}
\bibliography{references}

\renewcommand\thefigure{\thesection.\arabic{table}}
\setcounter{table}{0}

\appendix

\section{Ejection Models} \label{appendixA}

In Section \ref{models}, we assume that wide binaries break up when their host dense cores divide, producing two single stars that are still embedded.  Here, we consider an alternative picture, where one or both of the binary companions are ejected from the original dense core.  Such ejected stars have no dusty envelopes, and would not appear as Class 0 or I sources.  Instead, they would be identified as Class II or Class III objects, which we do not track.  The remaining stellar companion is still embedded, and counts in our populations.  Consider, for example, the w\_sb$^{\ast}$ model, the analogue to the w\_sb model described previously.  Its non-dimensional governing equations are:

\begin{subequations} \label{ejection_equation}
\begin{eqnarray}
dN_{0,w}/d\tau &=& (R\ \tOI) - N_{0,w}/(t_s/\tOI) -  \\
			& &  N_{0,w} - N_{0,w}/(t_b/\tOI)  \nonumber \\
dN_{0,t}/d\tau &=& N_{0,w}/(t_s/\tOI) - N_{0,t}   \\
d\NIw/d\tau &=& N_{0,w} - \NIw/(t_s/\tOI)  - \\
			& & \NIw/(t_b/\tOI) - \NIw/(\tI/\tOI) \nonumber  \\
d\NIt/d\tau &=& \NIw/(t_s/\tOI) + N_{0,t}  - \\
			& &	 \NIt/(\tI/\tOI)  \nonumber \\
dN_{0}/d\tau &=& N_{0,w}/(t_b/\tOI) - N_{0}   \\
d\NI/d\tau &=& \NIw/(t_b/\tOI) + N_{0} - \NI/(\tI/\tOI) \hspace{4mm}
\end{eqnarray}
\end{subequations}

As before, we apply a steady-state assumption to the six algebraic expressions for each model and solve for the set of parameters that give the lowest $\chi^2$.   Table \ref{modelTable2} lists the six unique single-star ejection models (e.g., the wi\_s and wti\_s models are not affected by changes in binary breakup) and the three possible double-star ejection models with their lowest $\chi^2$ values in the rightmost column.  The w\_sb$^{\ast}$ model produces the lowest error with a similar $\chi^2$ value as the analogous w\_sb model from Section \ref{models}.  Overall, the double-star ejection models produce inferior fits to the observations than the optimal w\_sb and  w\_sb$^{\ast}$ models. 

\begin{table}
\caption{Results for the Ejection Models}\label{modelTable2}
\begin{tabular}{lcccccc}%{lcccccccc}
\hline\hline
\multirow{2}{*}{Model} 	&  \multicolumn{3}{c}{Formed in the Cloud}	&  \multirow{2}{*}{shrink} & \multirow{2}{*}{break} & \multirow{2}{*}{$\chi^2$}  \\								%shrink			breakup
		&  wide	     	   & tight	         	& indiv.	      	   &		       		&	 			& \\
\hline
\multicolumn{7}{c}{Single-Star Ejection} \\
\hline
wt\_b$^{\ast}$	& $\checkmark$  &$\checkmark$  	&			   &		       		& $\checkmark$  	& 76\\ 
wti\_b$^{\ast}$	& $\checkmark$  & $\checkmark$  	& $\checkmark$  &		        		& $\checkmark$  	& 19 \\
w\_sb$^{\ast}$	& $\checkmark$  &		         	&			   & $\checkmark$  	& $\checkmark$  	& 1.5 \\ 
wt\_sb$^{\ast}$	& $\checkmark$  & $\checkmark$  	&			   & $\checkmark$  	& $\checkmark$  	& 76 \\ 
wi\_sb$^{\ast}$ 	& $\checkmark$  & 		         	& $\checkmark$  & $\checkmark$  	& $\checkmark$ 	& 3.1 \\
wti\_sb$^{\ast}$	& $\checkmark$  & $\checkmark$  	& $\checkmark$  & $\checkmark$  	& $\checkmark$  	& 19 \\
\hline
\multicolumn{7}{c}{Double-Star Ejection} \\
\hline
wti\_b$^{\ast\ast}$	& $\checkmark$  & $\checkmark$  	& $\checkmark$   & 			 	& $\checkmark$  	& 72 \\ 
wi\_sb$^{\ast\ast}$ 	& $\checkmark$  & 		         	& $\checkmark$  & $\checkmark$  	& $\checkmark$ 	& 3.1 \\
wti\_sb$^{\ast\ast}$	& $\checkmark$  & $\checkmark$  	& $\checkmark$  & $\checkmark$  	& $\checkmark$  	& 72 \\
\hline
\end{tabular}
\end{table}

The non-dimensional parameter values for the w\_sb$^{\ast}$ model are $R\ \tOI = 24.67$, $t_s/\tOI = 1.47$, $t_b/\tOI = 0.37$, and $\tI/\tOI = 1.52$.   The resulting stellar birth rate is 247 Myr$^{-1}$,  again assuming $\tOI = 0.2$ Myr.  This rate is $\sim 28$\%\ higher than the birth rate estimated by \citet{Evans09}.

\section{Individual Binary-Core Systems} \label{appendix}

Images of all 24 embedded binary systems in Perseus.  Each image shows a $63\arcsec\ \times 63\arcsec$ cutout around each core.  These cutouts have a  corresponding physical scale of $\sim 0.7\ \mbox{pc} \times 0.7\ \mbox{pc}$, for a distance of 235 pc to Perseus \citep{Hirota08}.  The blue contour on each image corresponds to 50\%\ of the peak core flux.  Each map has a spatial resolution of 14.6\arcsec.  Embedded sources detected in the VANDAM survey \citep{Tobin16} are labeled in white.  The physical properties of the cores from the \emph{getsources} extractions are given in Table \ref{coreTable}.

\onecolumn

\renewcommand\thefigure{\thesection.\arabic{figure}}
\setcounter{figure}{0}

\begin{figure}
\centering
\subfloat{\label{fig:1}\includegraphics[width=0.49\textwidth,trim=1pt 1pt 1pt 1pt,clip=true]{plotWest47.eps}}
\subfloat{\label{fig:2}\includegraphics[width=0.49\textwidth,trim=1pt 1pt 1pt 1pt,clip=true]{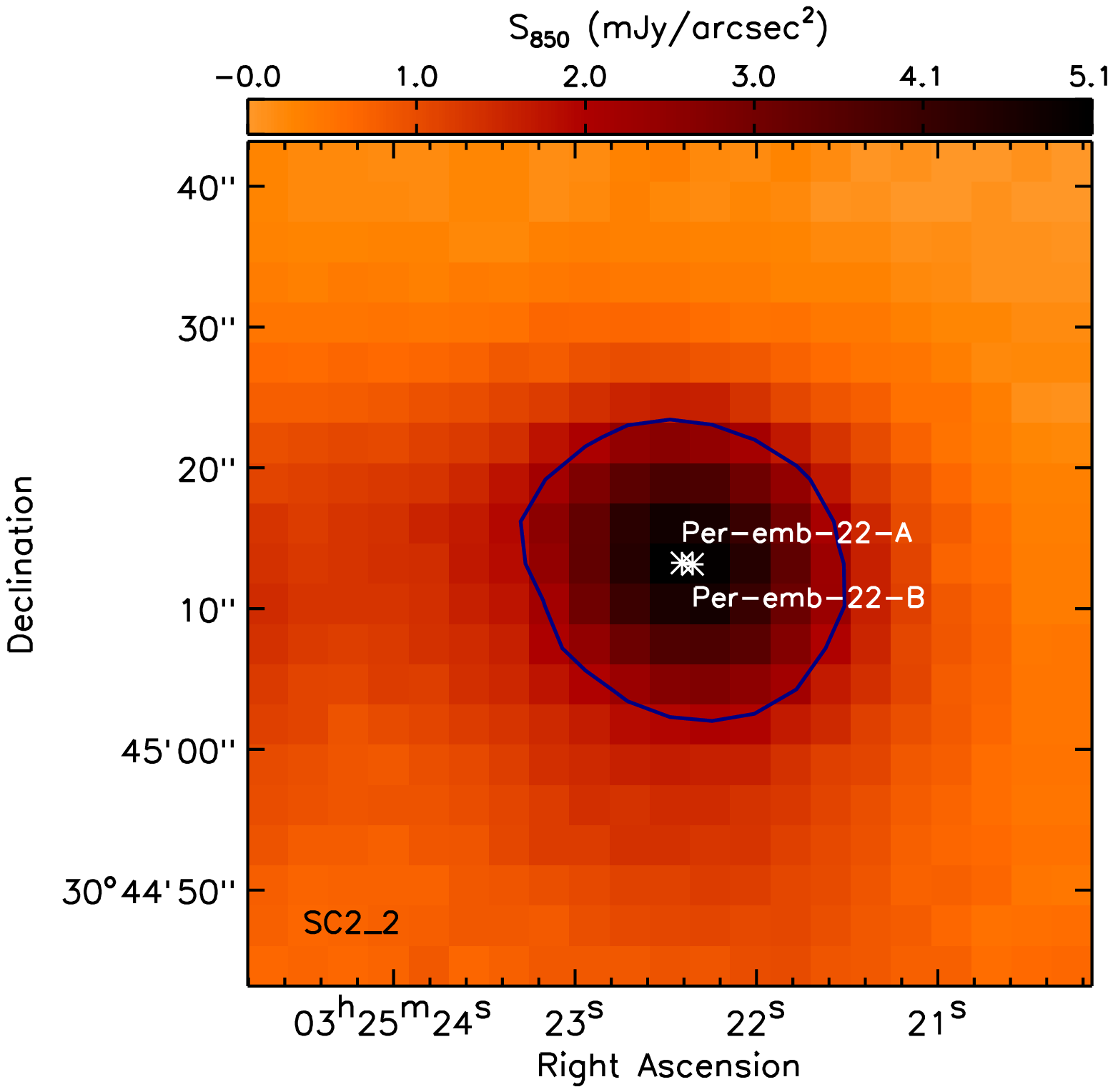}}
\quad
\subfloat{\label{fig:3}\includegraphics[width=0.49\textwidth,trim=1pt 1pt 1pt 1pt,clip=true]{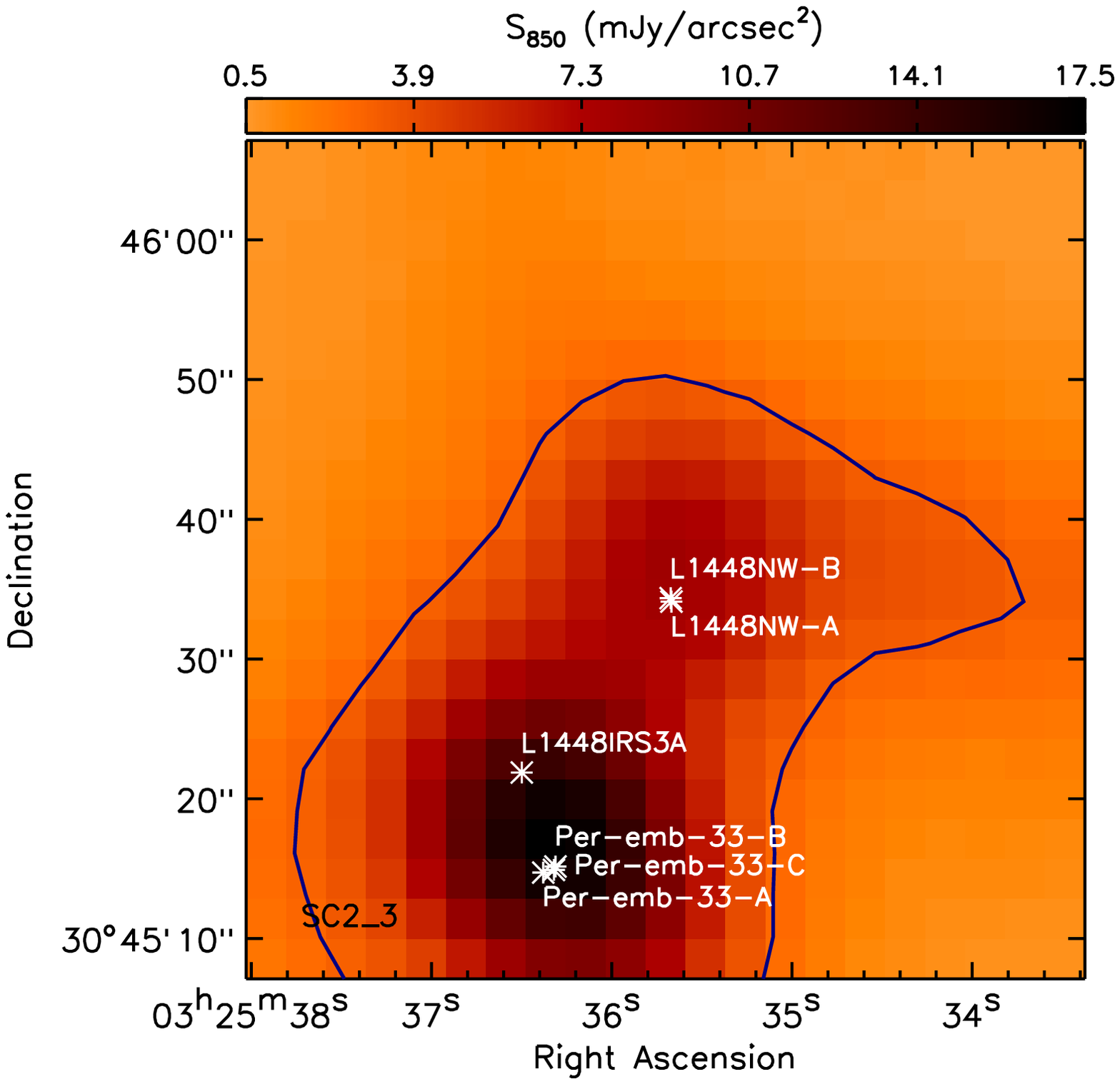}}
\subfloat{\label{fig:4}\includegraphics[width=0.49\textwidth,trim=1pt 1pt 1pt 1pt,clip=true]{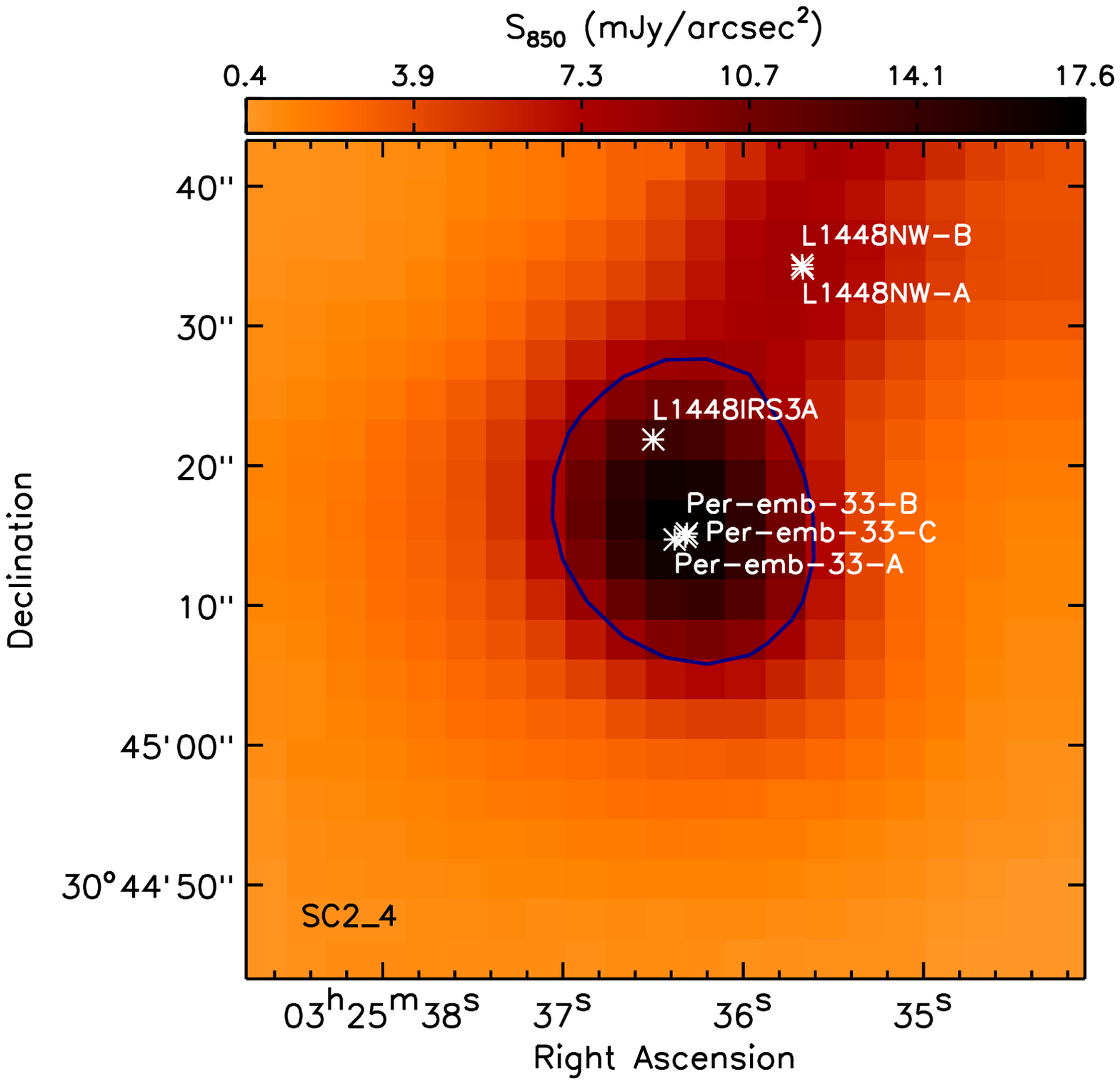}}
\quad
\caption{SCUBA-2 850 \um\ maps of the host cores for all the embedded multiples in our sample.  Each map is $\sim 1$\arcmin\ on a side, centered on each core (see Table \ref{coreTable}).  VLA-identified sources are labeled on each map with white stars and labels.  Blue contours corresponds to emission at 50\%\ of the peak core flux.  Core identifications are given in the lower-left corner. }
\label{fig:allfeature1}
\end{figure}
\begin{figure}
%\ContinuedFloat
\centering
\subfloat{\label{fig:7}\includegraphics[width=0.49\textwidth,trim=1pt 1pt 1pt 1pt,clip=true]{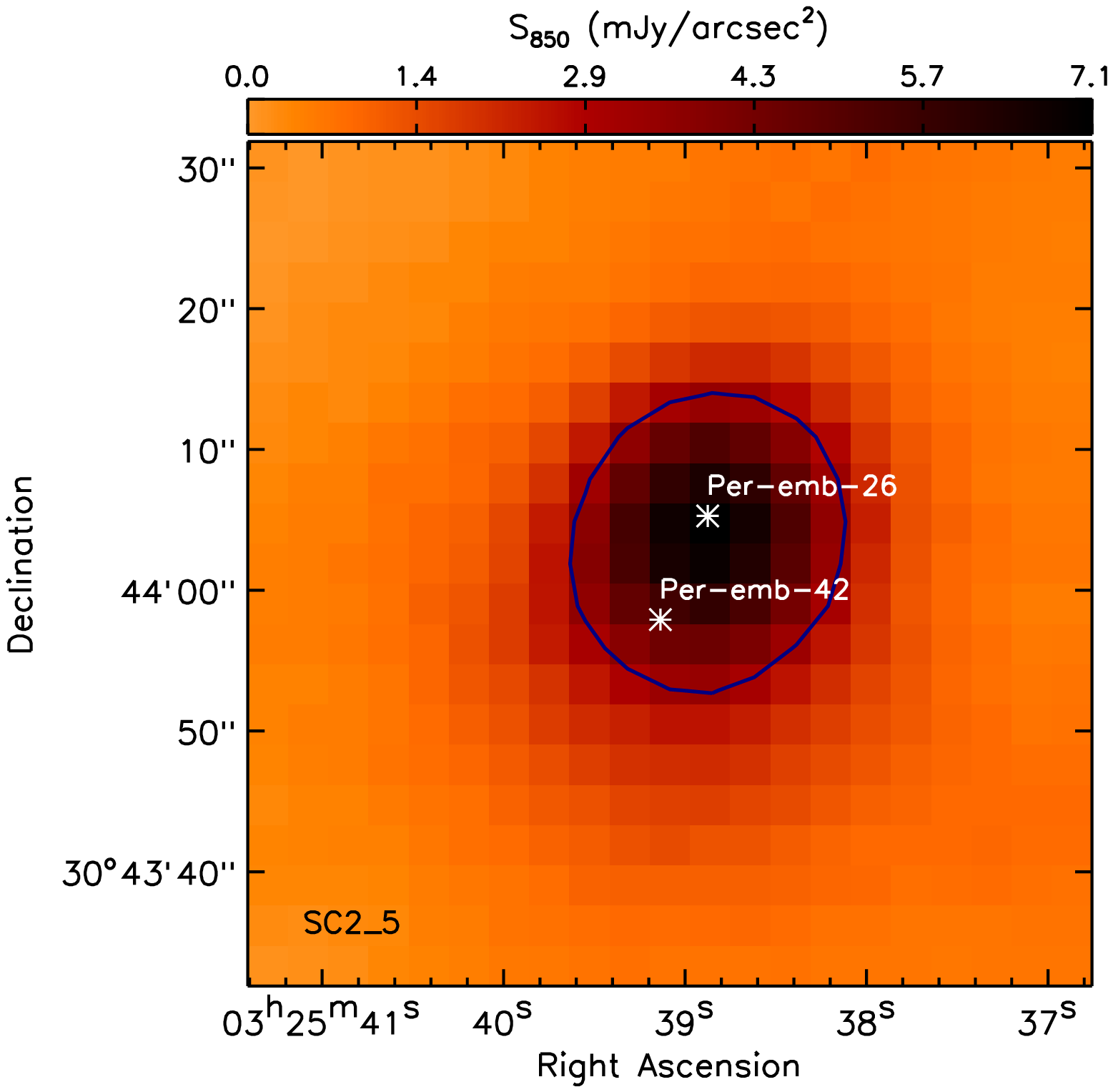}}
\subfloat{\label{fig:8}\includegraphics[width=0.49\textwidth,trim=1pt 1pt 1pt 1pt,clip=true]{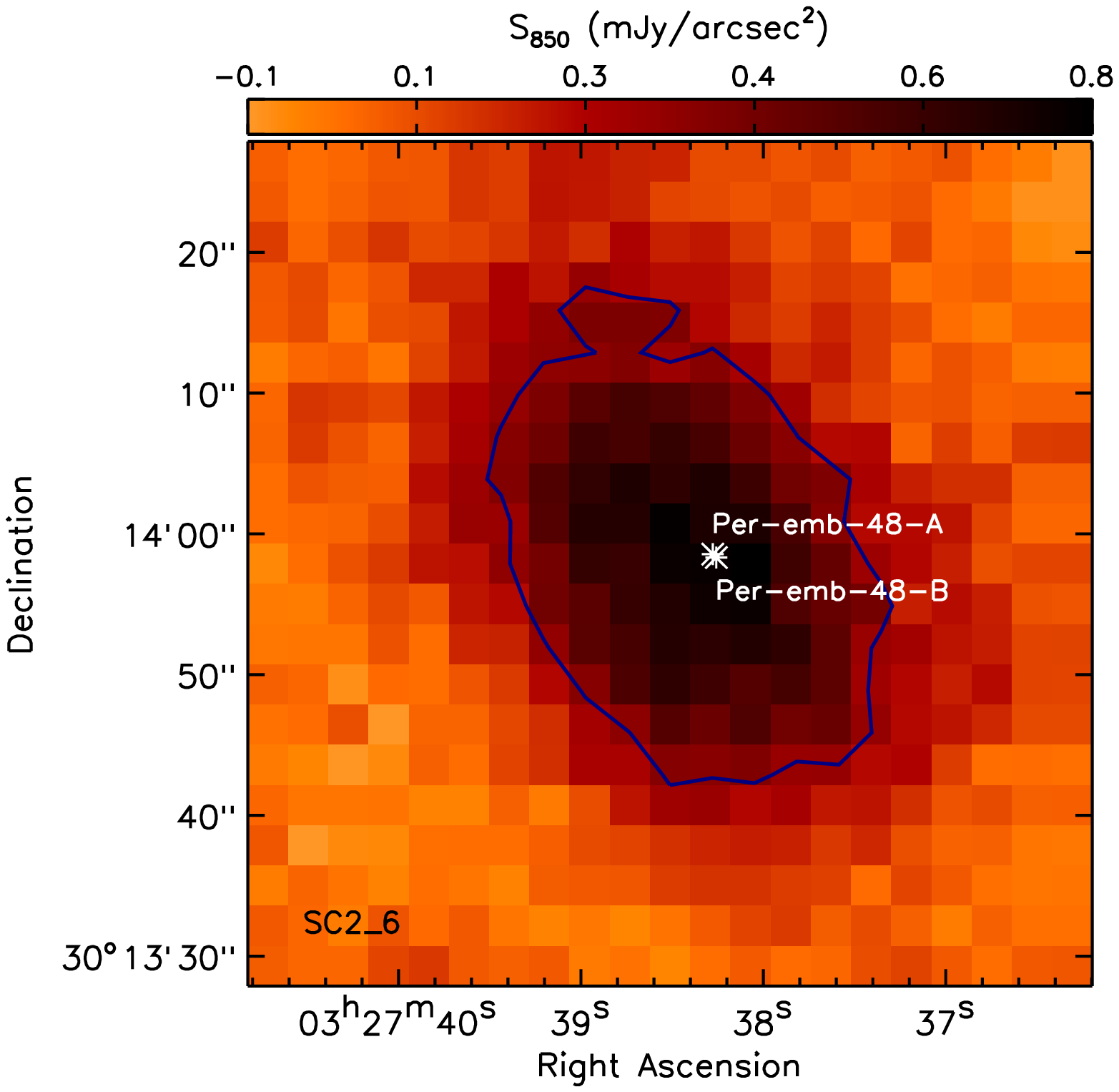}}
\quad
\subfloat{\label{fig:9}\includegraphics[width=0.49\textwidth,trim=1pt 1pt 1pt 1pt,clip=true]{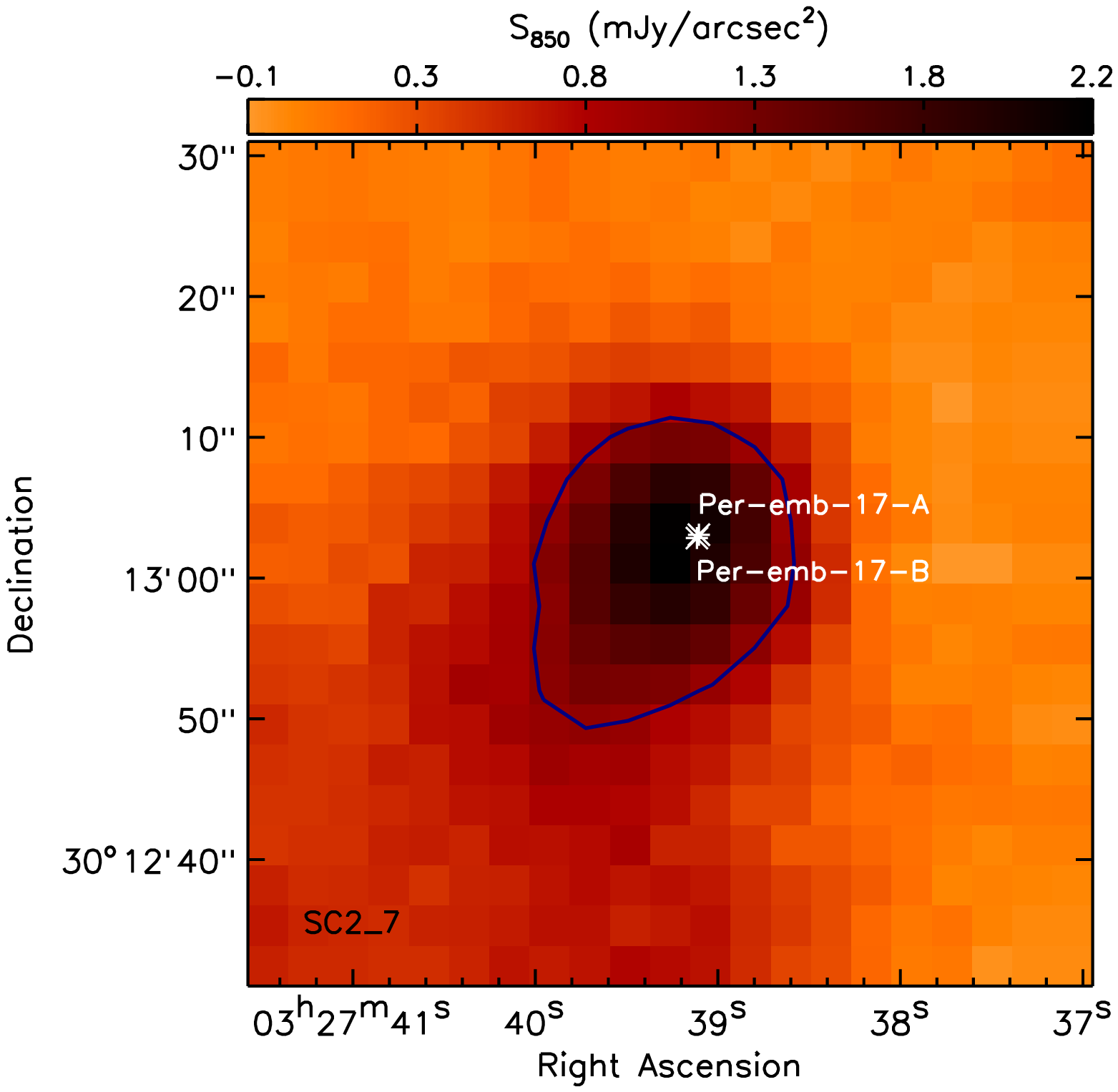}}
\subfloat{\label{fig:10}\includegraphics[width=0.49\textwidth,trim=1pt 1pt 1pt 1pt,clip=true]{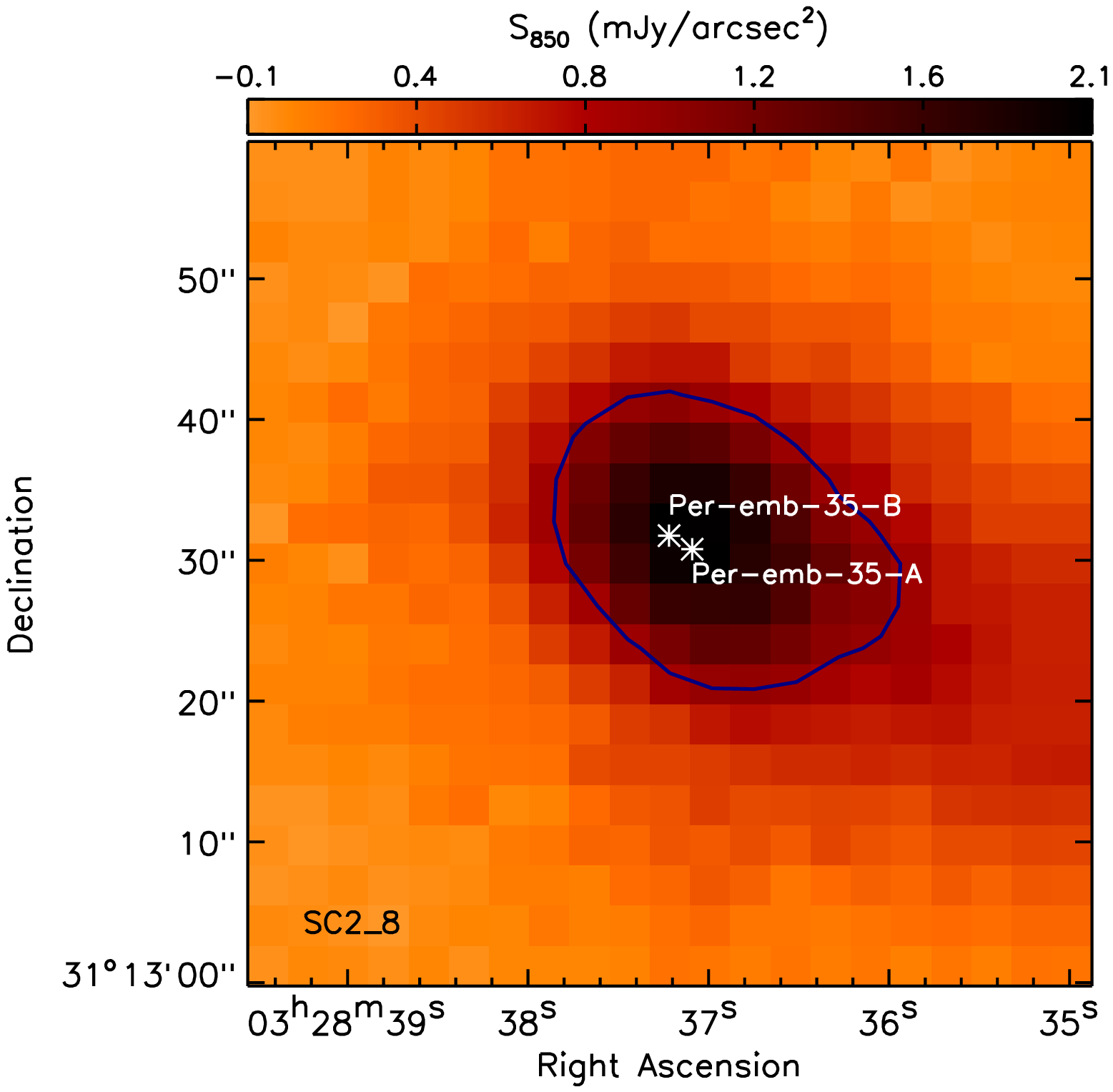}}
\quad
\contcaption{}
\label{fig:allfeature2}
\end{figure}
\begin{figure}
%\ContinuedFloat
\centering
\subfloat{\label{fig:13}\includegraphics[width=0.49\textwidth,trim=1pt 1pt 1pt 1pt,clip=true]{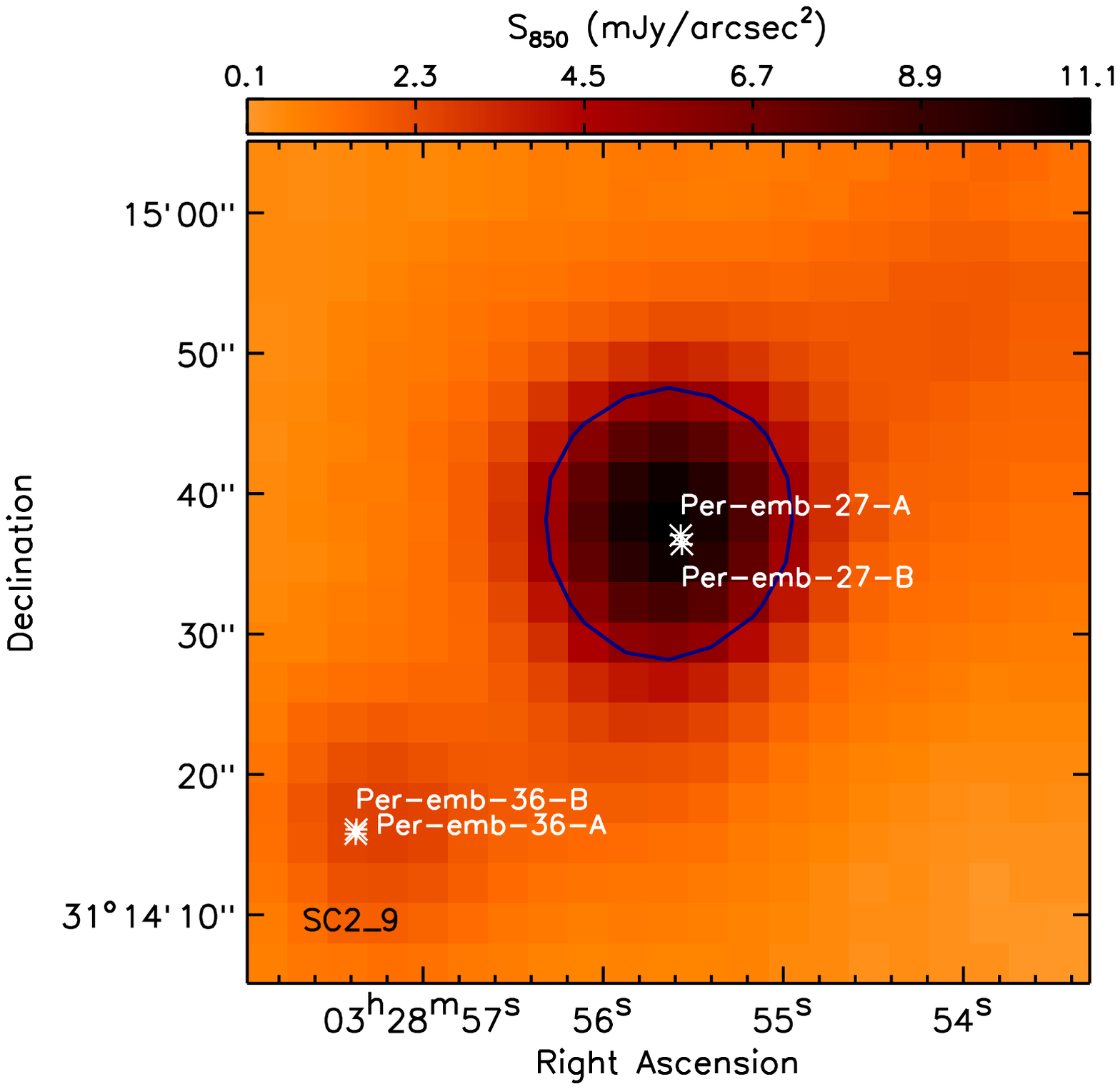}}
\subfloat{\label{fig:14}\includegraphics[width=0.49\textwidth,trim=1pt 1pt 1pt 1pt,clip=true]{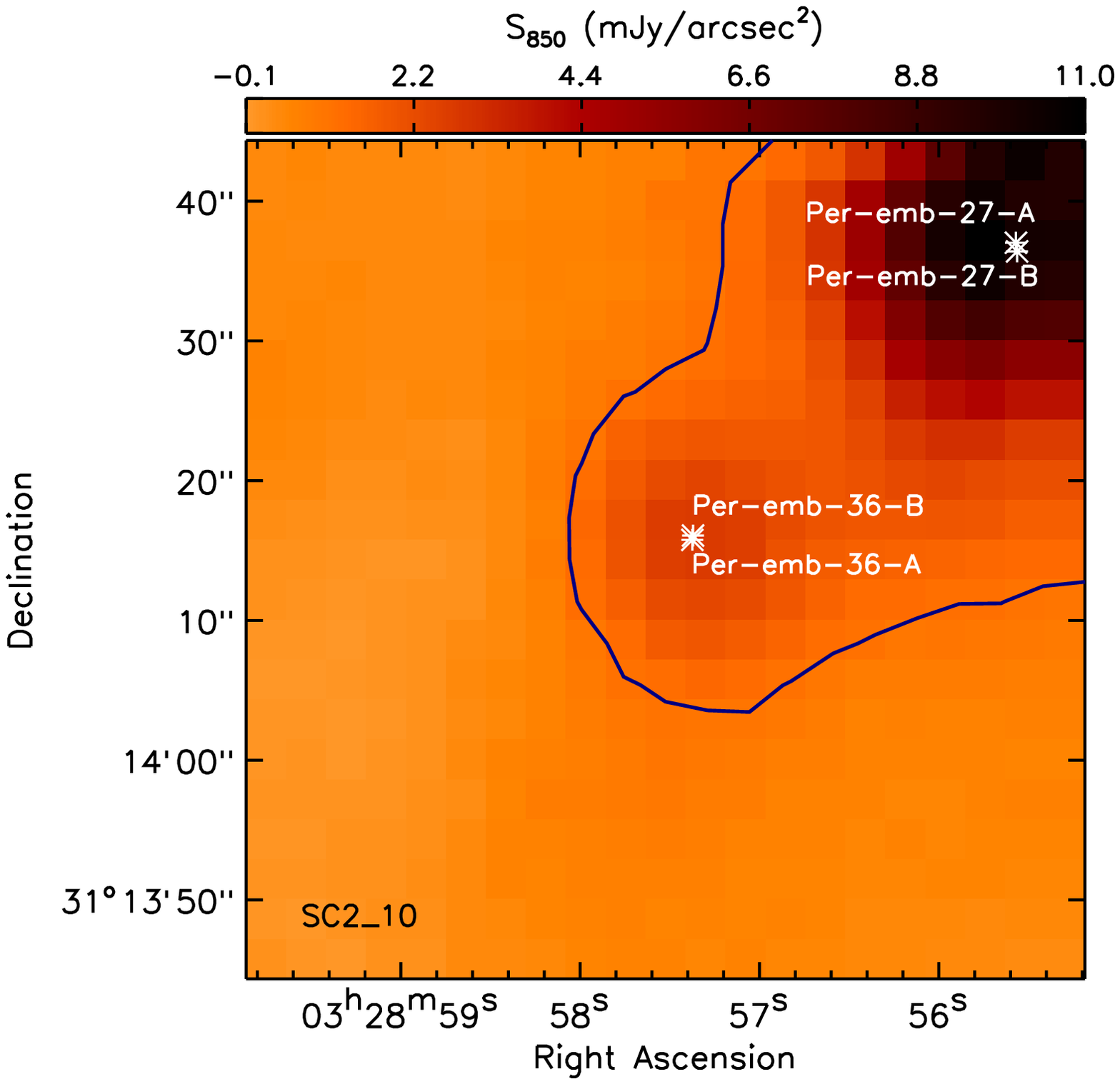}}
\quad
\subfloat{\label{fig:15}\includegraphics[width=0.49\textwidth,trim=1pt 1pt 1pt 1pt,clip=true]{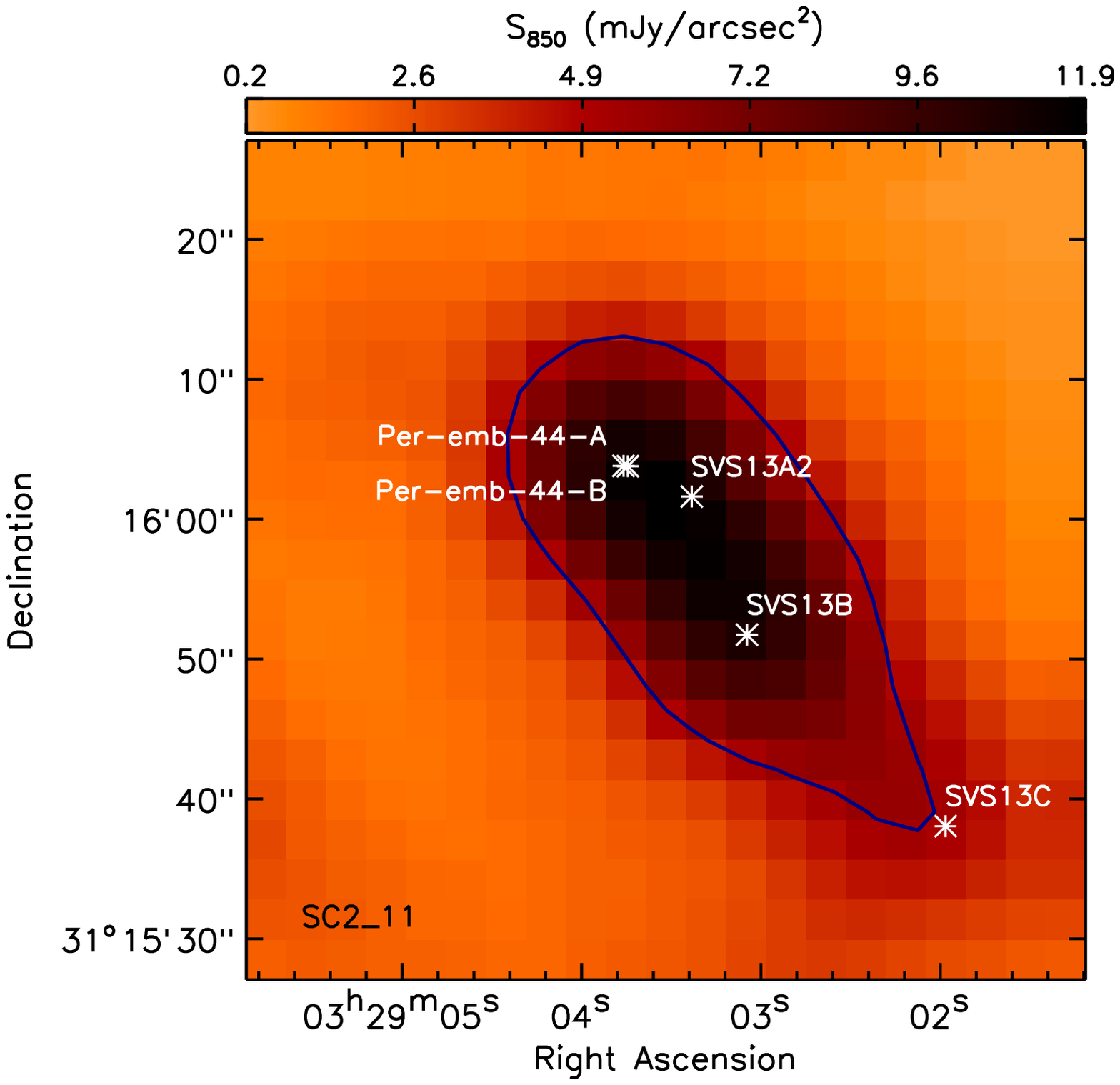}}
\subfloat{\label{fig:16}\includegraphics[width=0.49\textwidth,trim=1pt 1pt 1pt 1pt,clip=true]{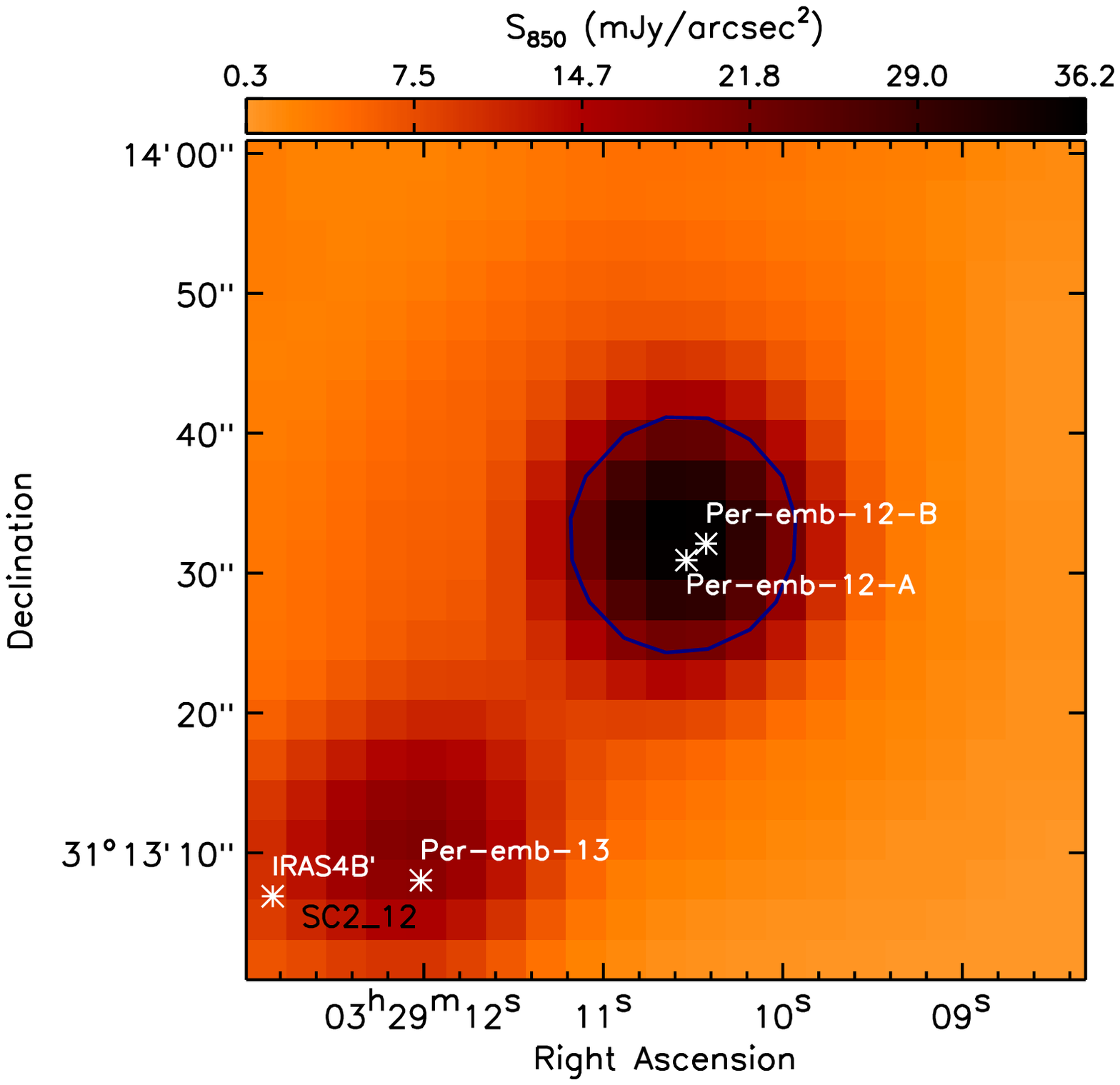}}
\quad
\contcaption{}
\label{fig:allfeature3}
\end{figure}
\begin{figure}
%\ContinuedFloat
\centering
\subfloat{\label{fig:19}\includegraphics[width=0.49\textwidth,trim=1pt 1pt 1pt 1pt,clip=true]{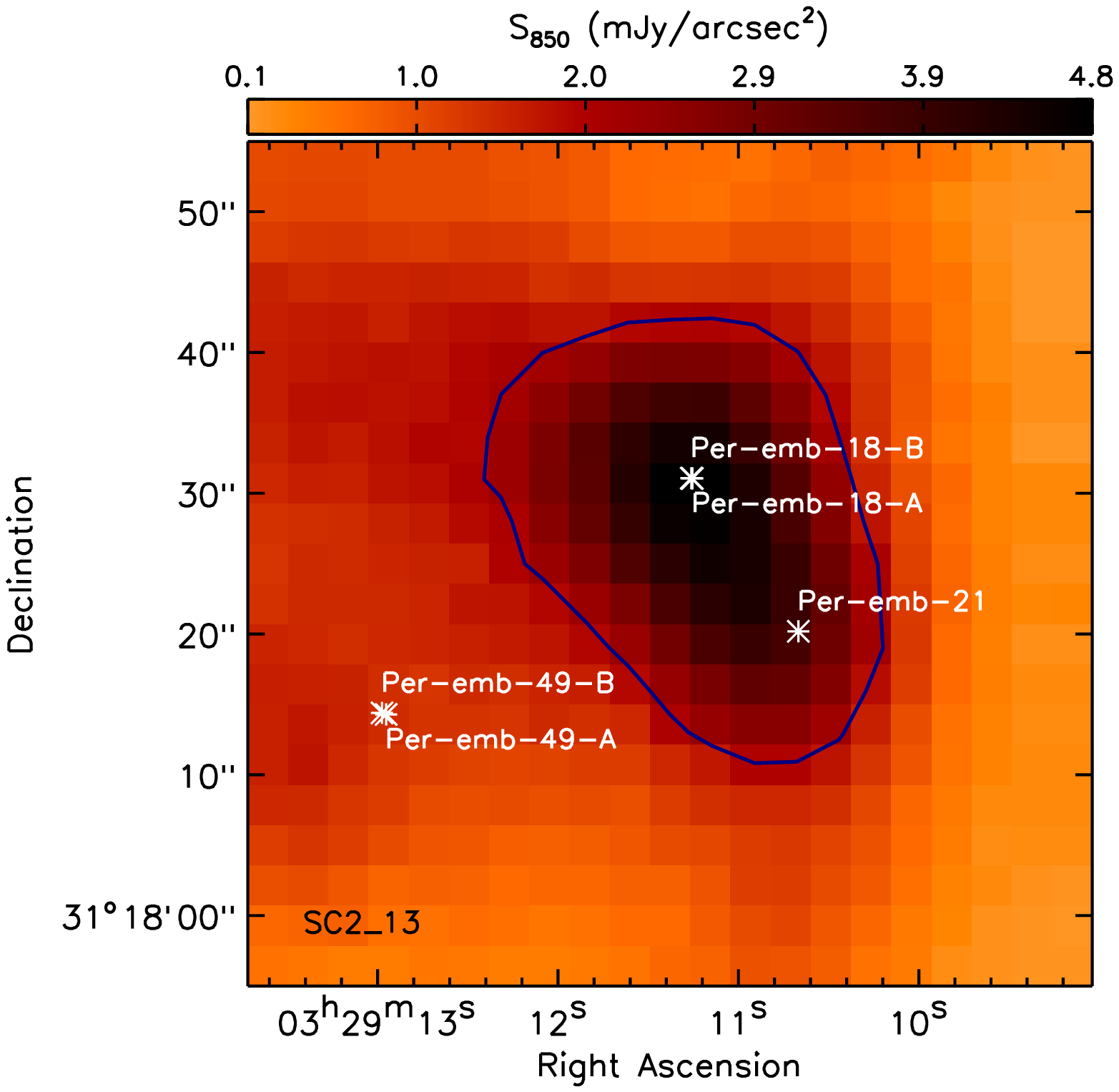}}
\subfloat{\label{fig:20}\includegraphics[width=0.49\textwidth,trim=1pt 1pt 1pt 1pt,clip=true]{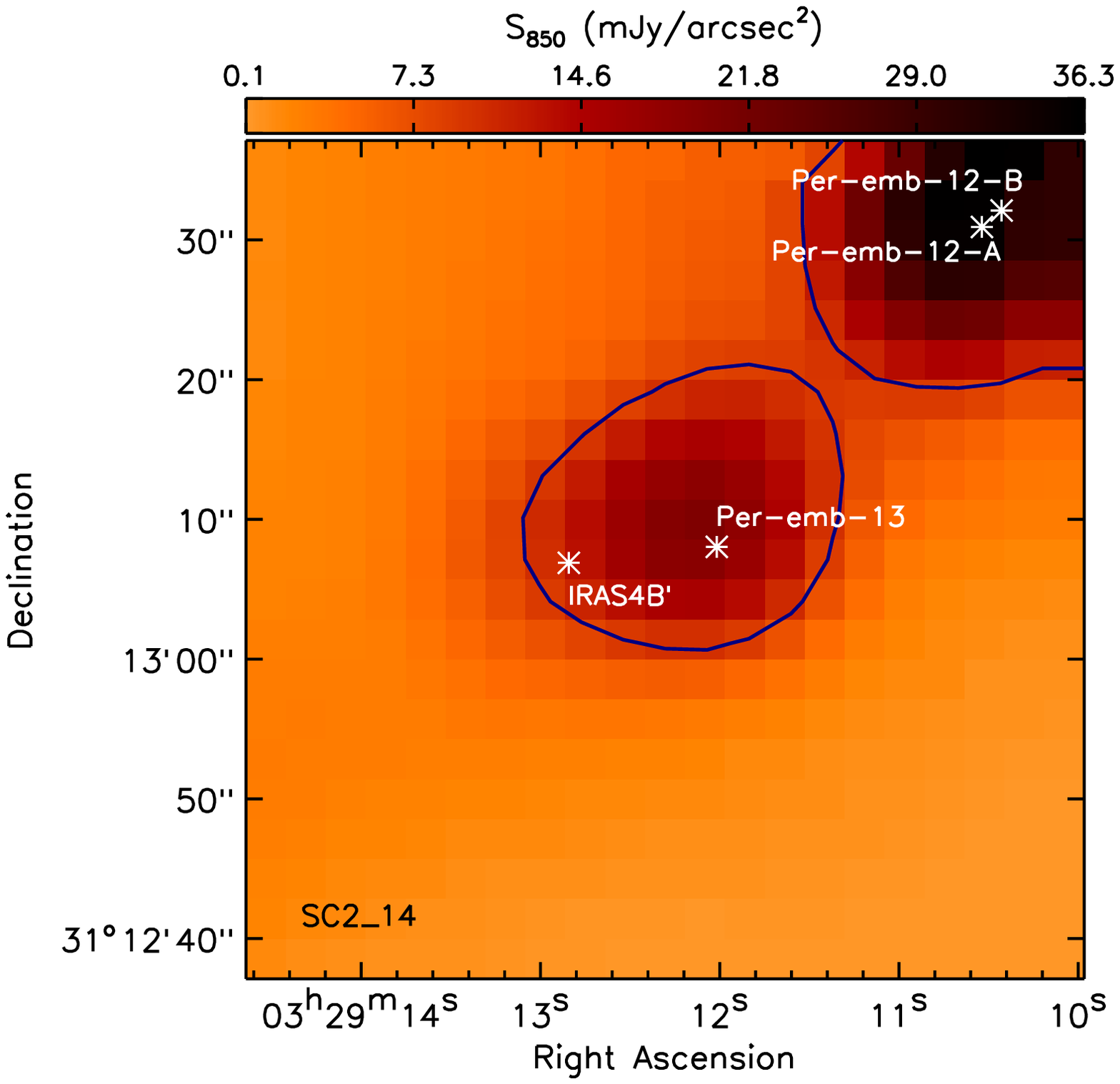}}
\quad
\subfloat{\label{fig:21}\includegraphics[width=0.49\textwidth,trim=1pt 1pt 1pt 1pt,clip=true]{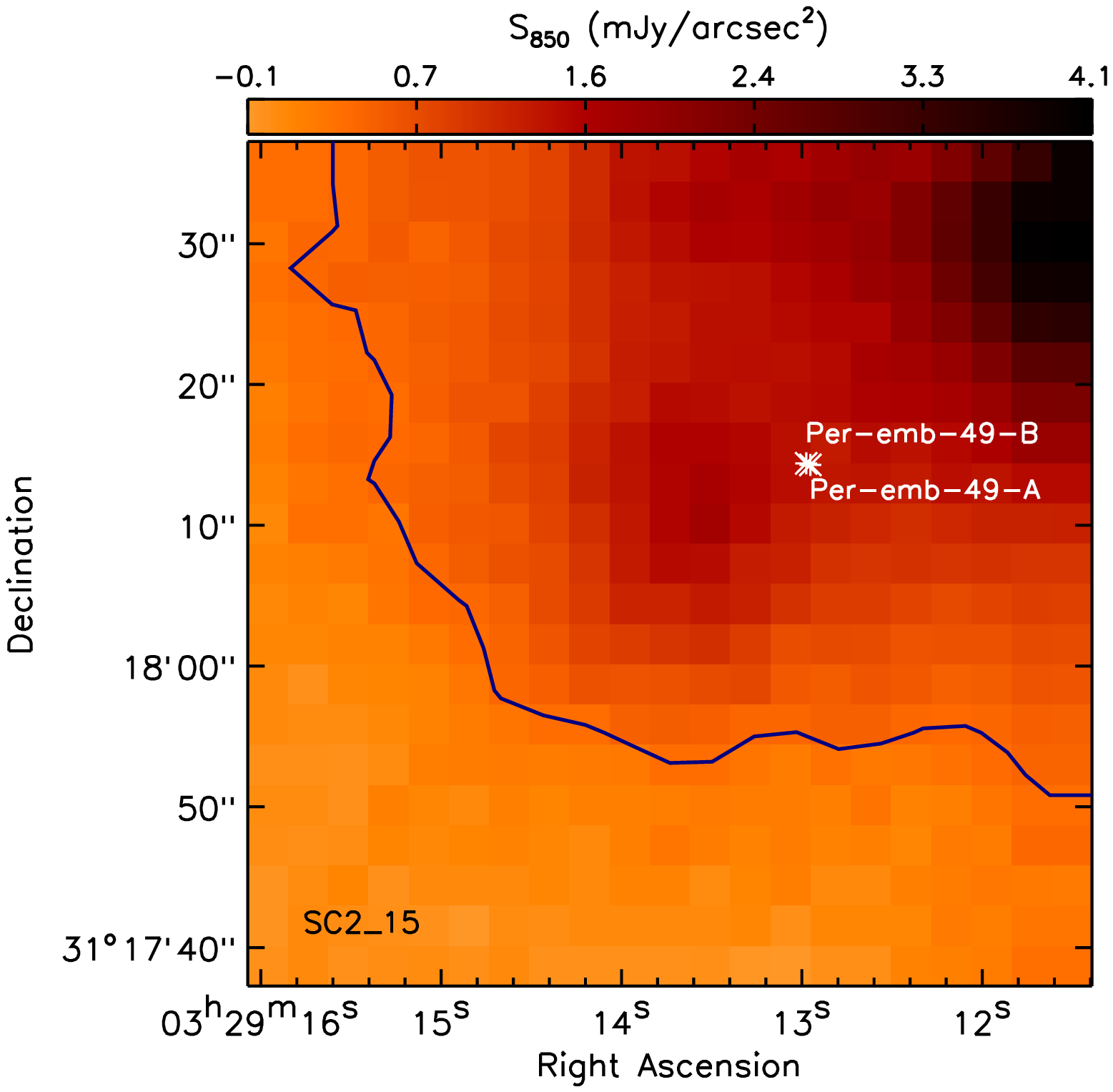}}
\subfloat{\label{fig:22}\includegraphics[width=0.49\textwidth,trim=1pt 1pt 1pt 1pt,clip=true]{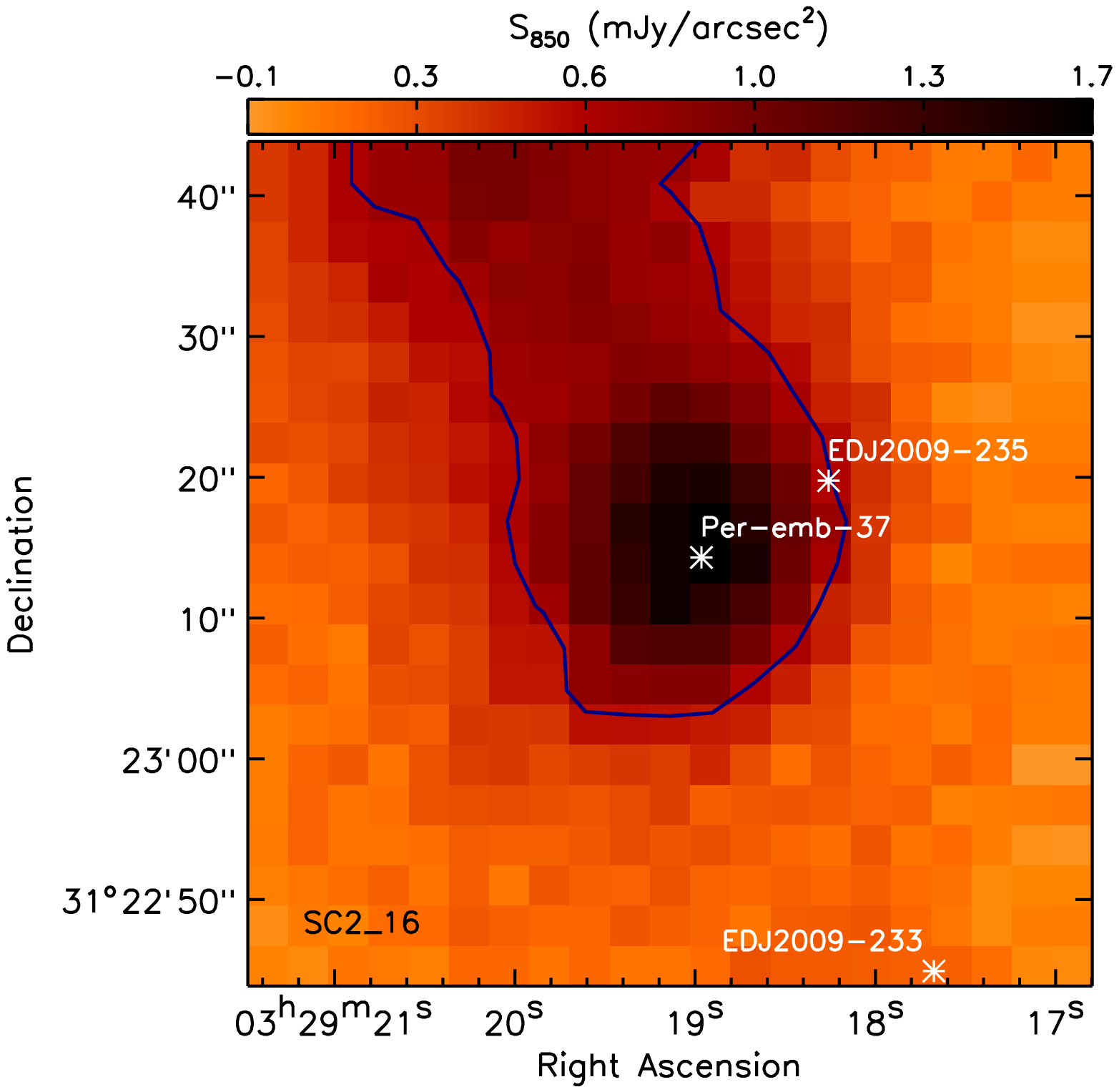}}
\quad
\contcaption{}
\label{fig:allfeature4}
\end{figure}
\begin{figure}
%\ContinuedFloat
\centering
\subfloat{\label{fig:25}\includegraphics[width=0.49\textwidth,trim=1pt 1pt 1pt 1pt,clip=true]{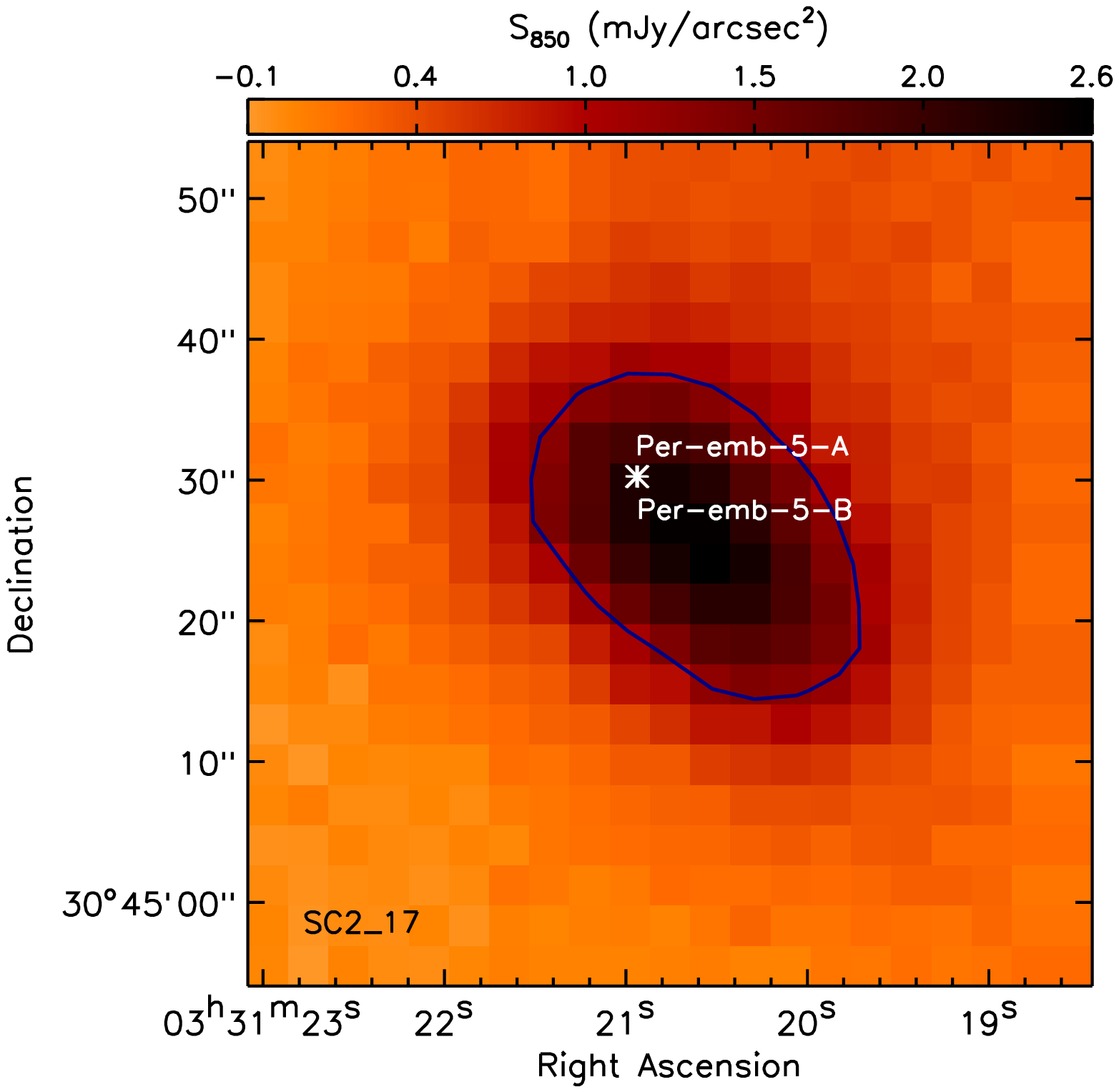}}
\subfloat{\label{fig:26}\includegraphics[width=0.49\textwidth,trim=1pt 1pt 1pt 1pt,clip=true]{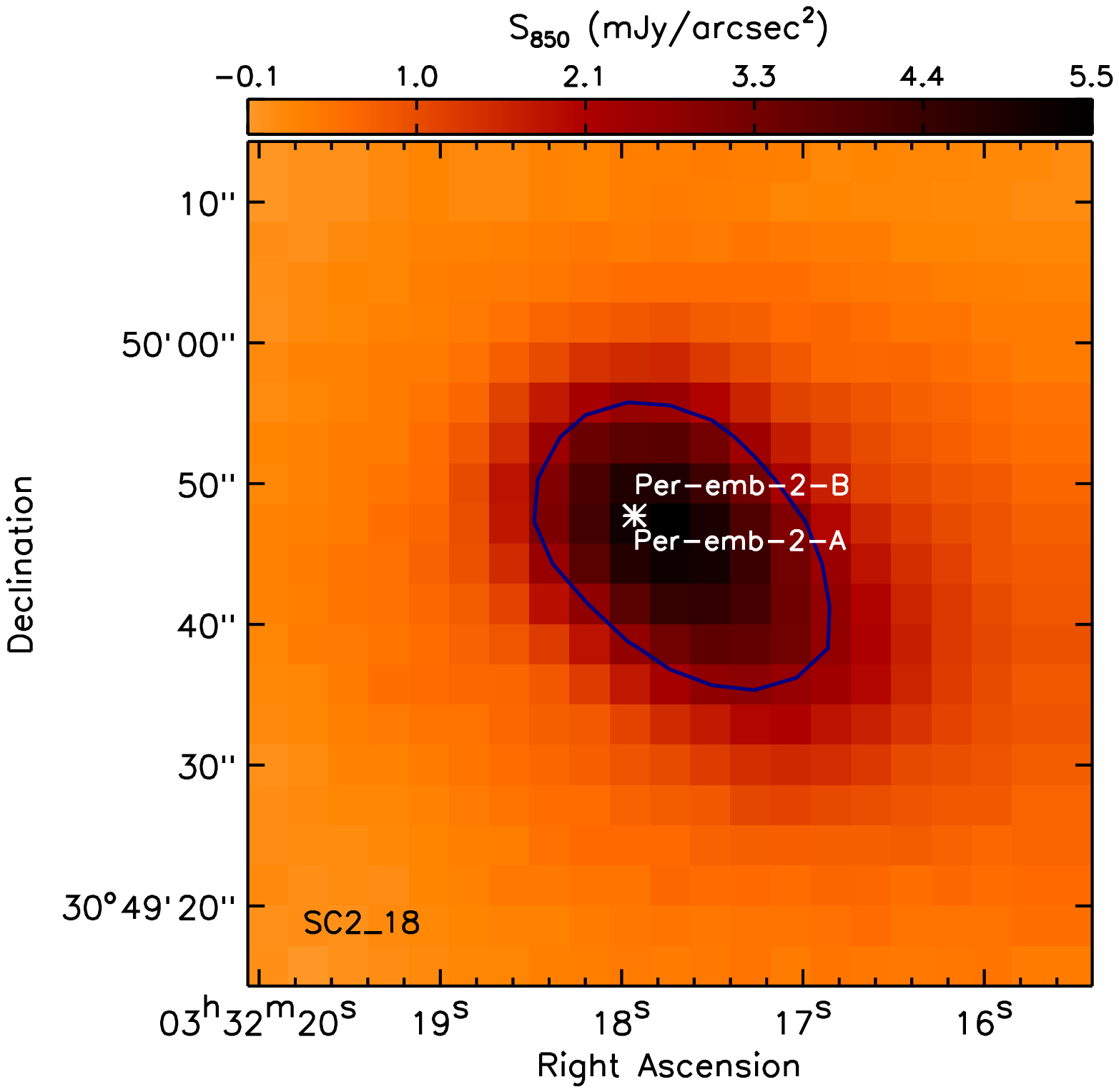}}
\quad
\subfloat{\label{fig:27}\includegraphics[width=0.49\textwidth,trim=1pt 1pt 1pt 1pt,clip=true]{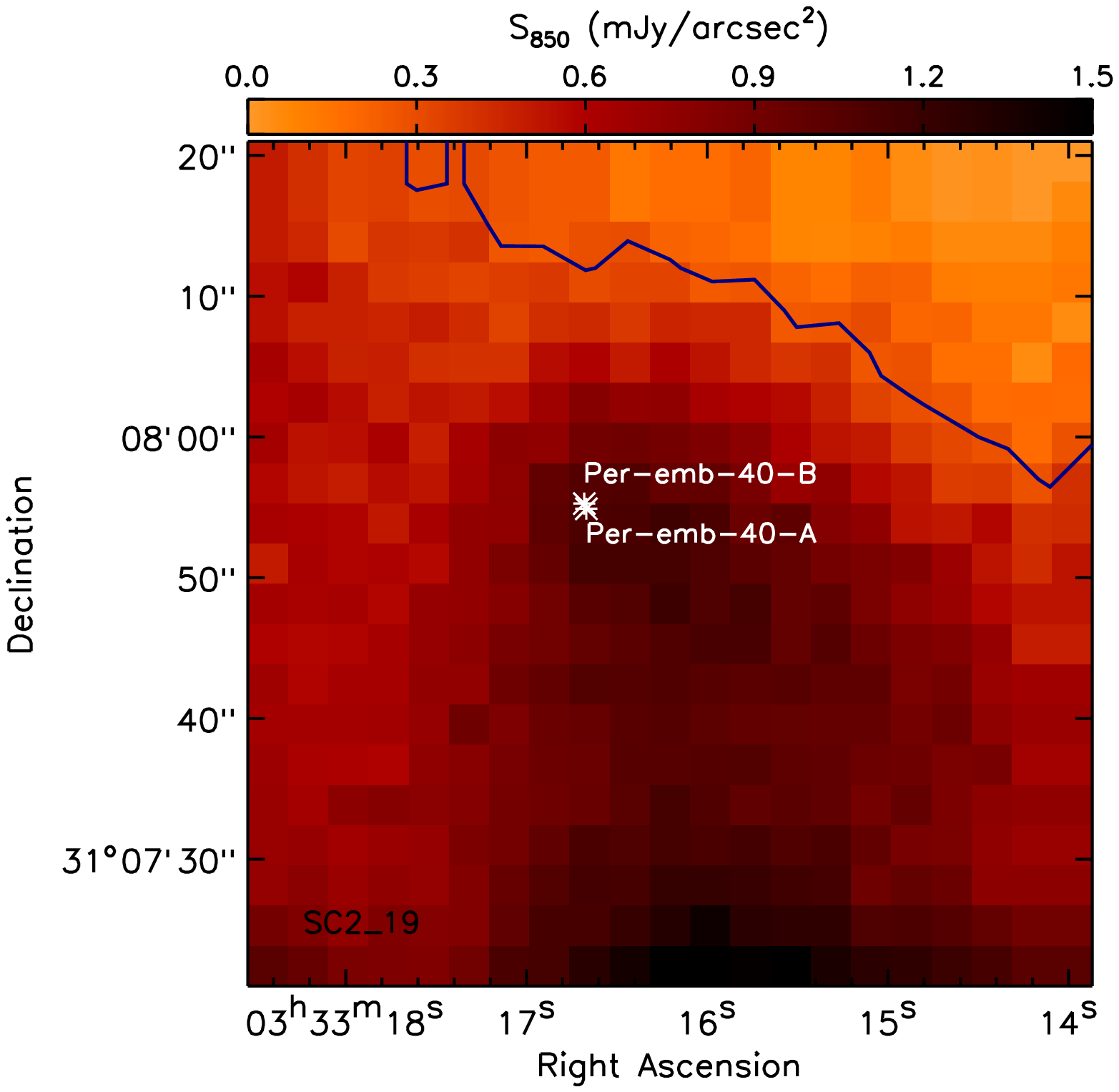}}
\subfloat{\label{fig:28}\includegraphics[width=0.49\textwidth,trim=1pt 1pt 1pt 1pt,clip=true]{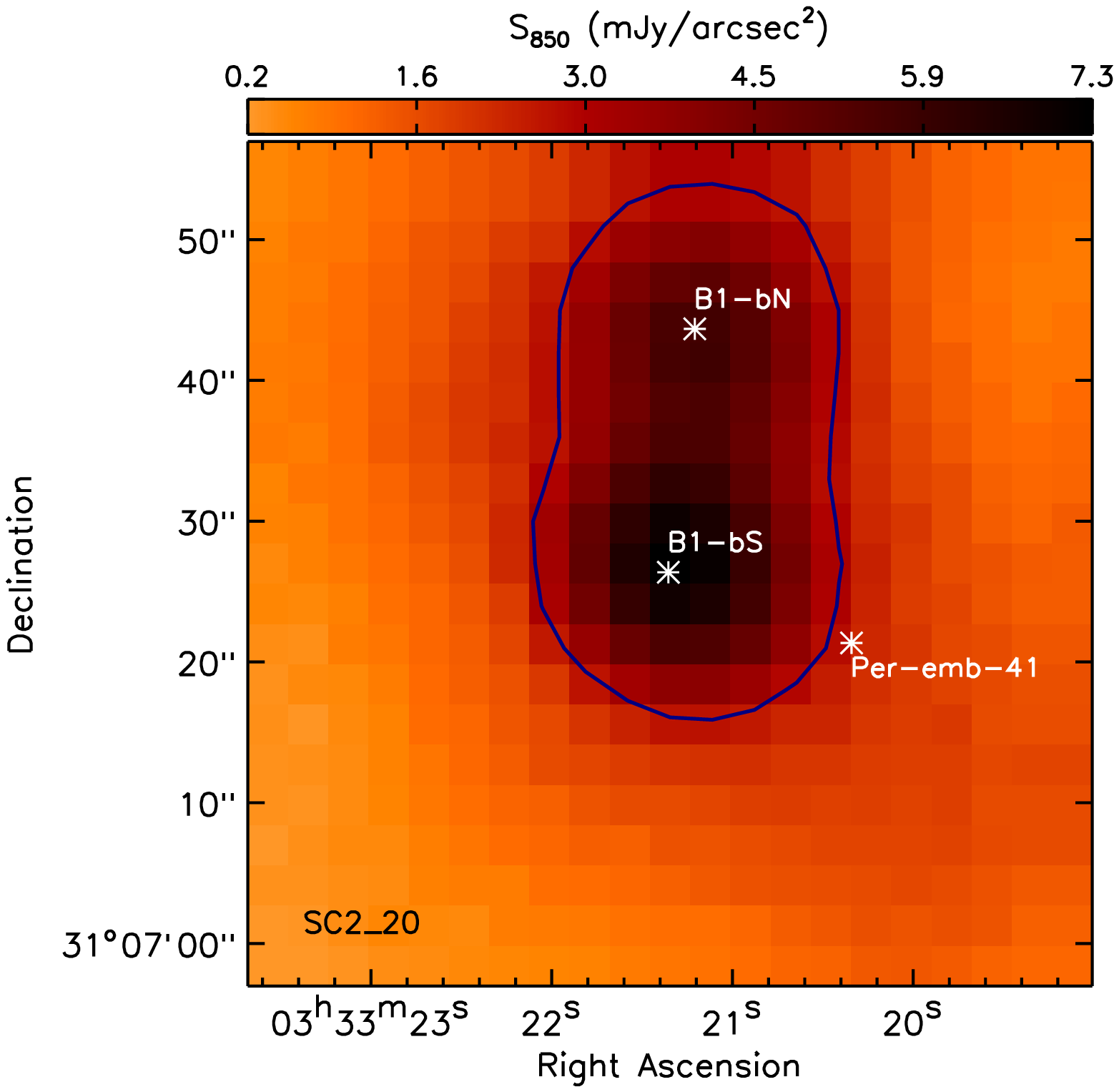}}
\quad
\contcaption{}
\label{fig:allfeature5}
\end{figure}
\begin{figure}
%\ContinuedFloat
\centering
\subfloat{\includegraphics[width=0.49\textwidth,trim=1pt 1pt 1pt 1pt,clip=true]{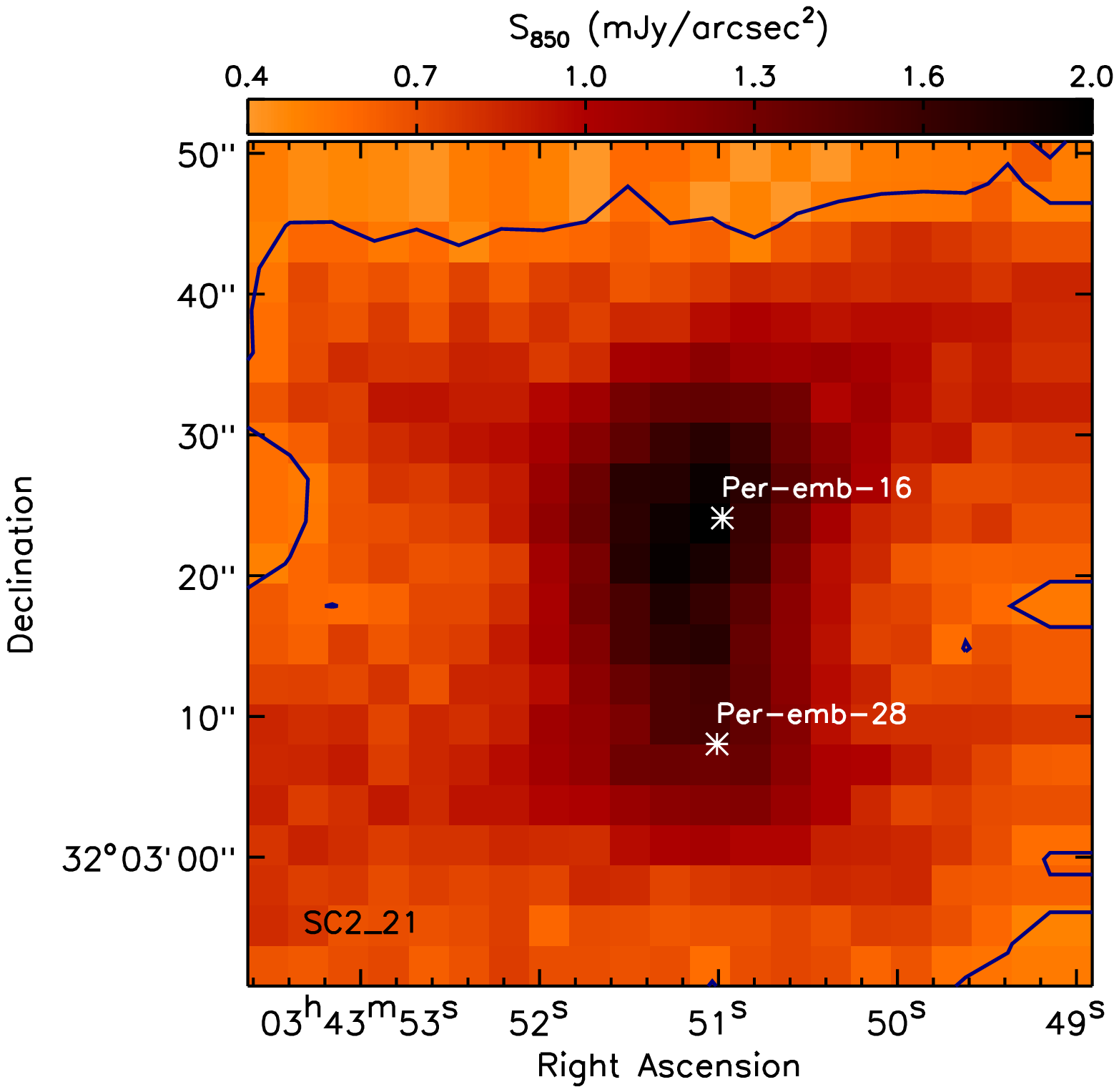}}
\subfloat{\includegraphics[width=0.49\textwidth,trim=1pt 1pt 1pt 1pt,clip=true]{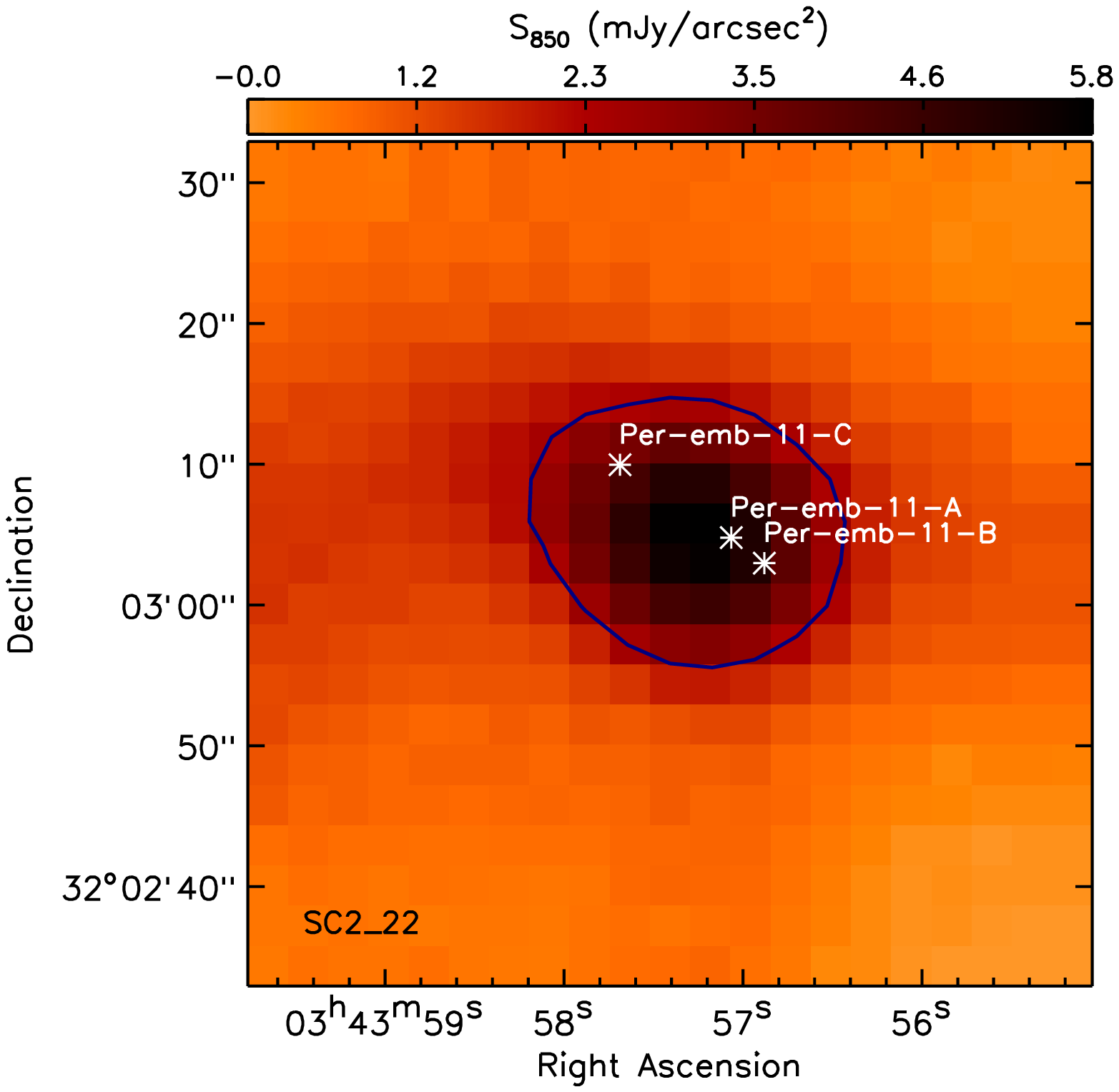}}
\quad
\subfloat{\includegraphics[width=0.49\textwidth,trim=1pt 1pt 1pt 1pt,clip=true]{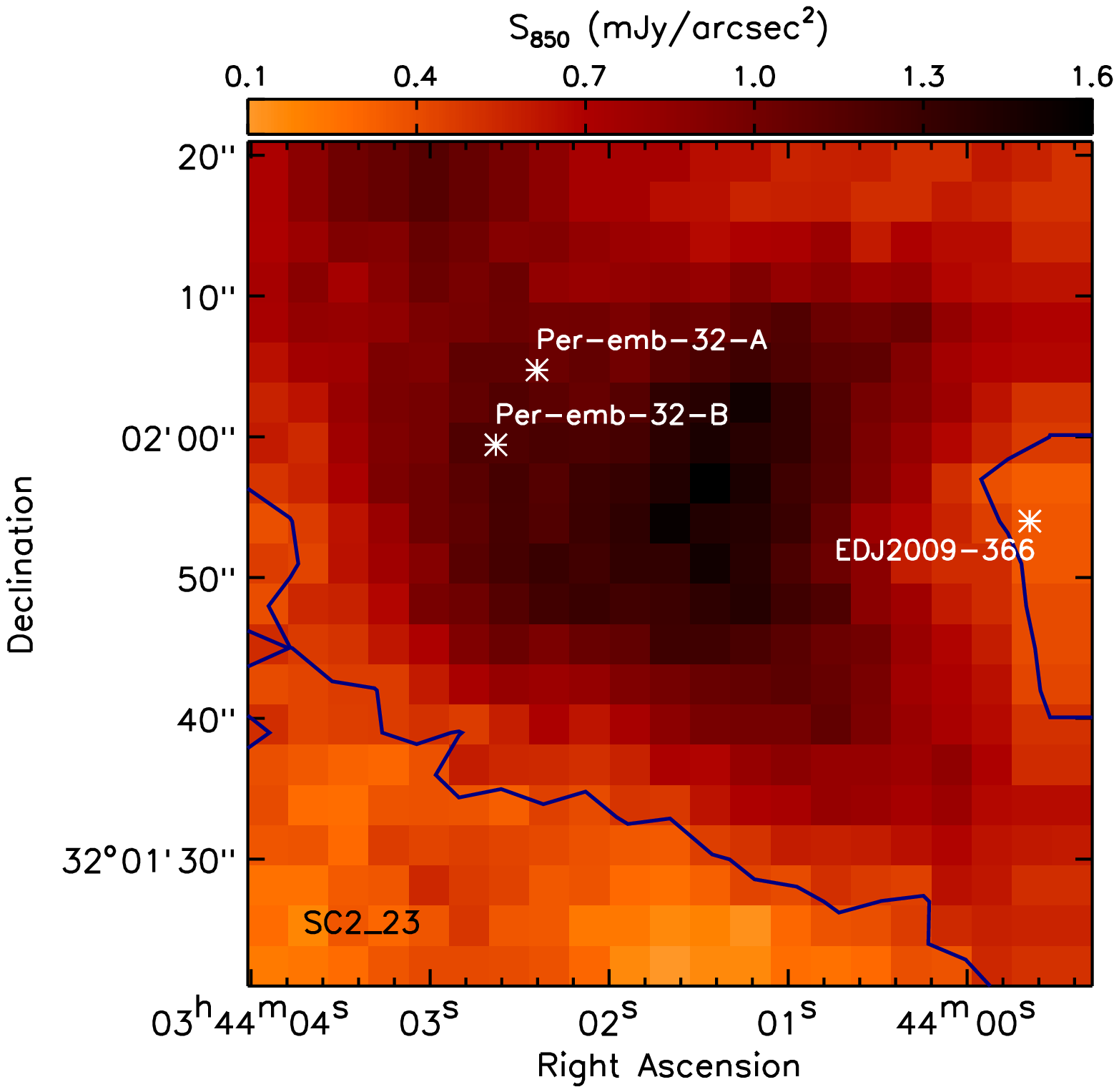}}
\subfloat{\includegraphics[width=0.49\textwidth,trim=1pt 1pt 1pt 1pt,clip=true]{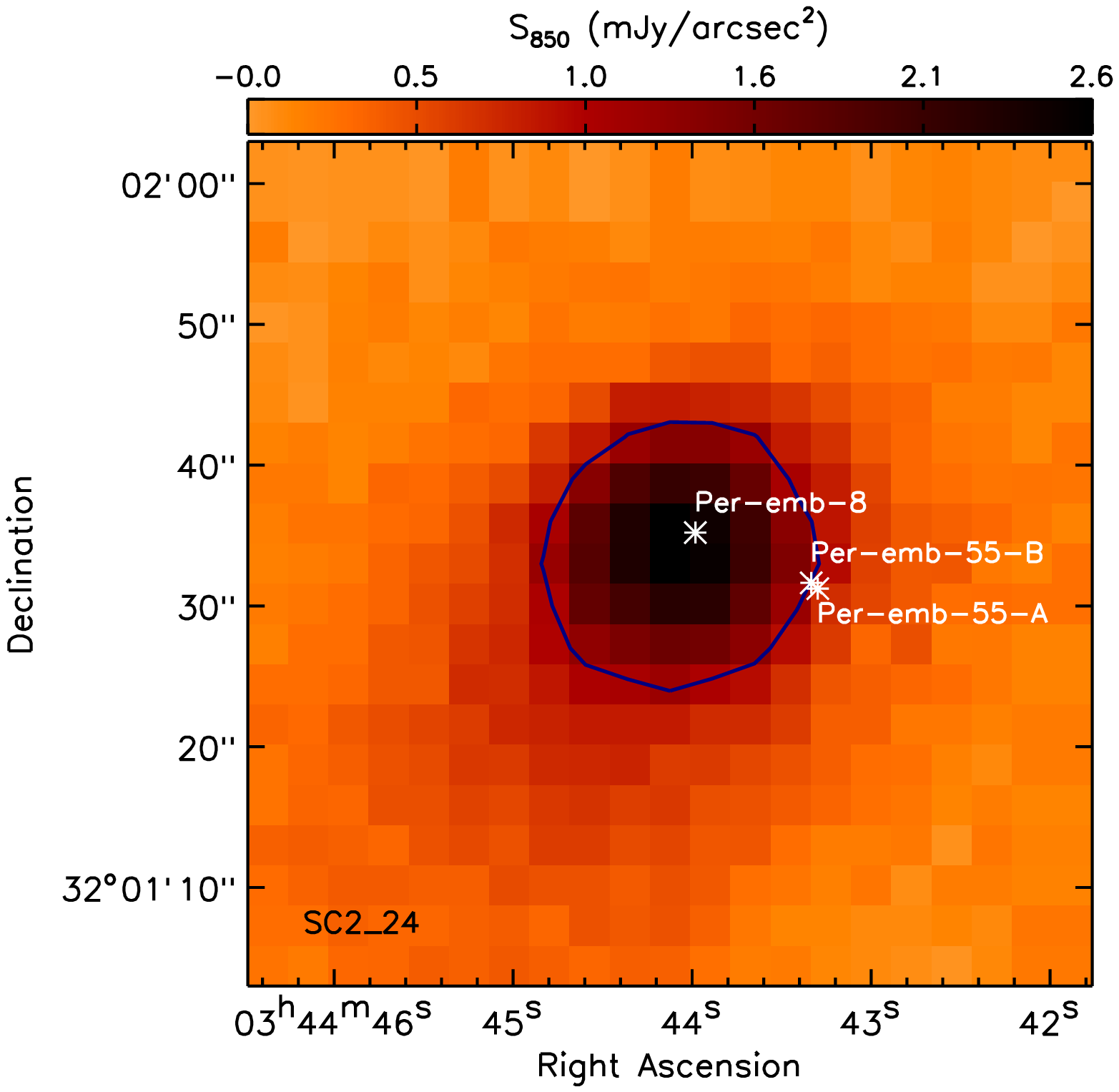}}
\quad
\contcaption{}
\label{fig:allfeature7}
\end{figure}

\bsp
\label{lastpage}

\end{document}